**Efficient purely organic phosphorescent emitters for programmable luminescent tags: from building blocks to donor-acceptor-donor structures**

Uliana Tsiko[1], Sebastian Kaiser[1], Jannis Fidelius[2], Tim Achenbach[1,3], Jan J. Weigand[2], Sebastian Reineke[1], and Karl Sebastian Schellhammer[1]*

[1]Dresden Integrated Center for Applied Physics and Photonic Materials (IAPP) and Institute of Applied Physics (IAP), Technische Universität Dresden, Hermann-Krone-Bau, Nöthnitzer Str. 61, 01187 Dresden, Germany

[2]Chair of Inorganic Molecular Chemistry, Faculty of Chemistry and Food Chemistry, Technische Universität Dresden, 01069 Dresden, Germany

[3]PRUUVE GmbH, 01187 Dresden, Germany

Correspondence e-mail: sebastian.schellhammer@tu-dresden.de

## Abstract

Purely organic room-temperature phosphorescence (RTP) emitters are key components of programmable luminescent tags (PLTs), photonic devices for rewritable information storage and UV dosimetry. In this work, we systematically explore the design space of donor-acceptor and donor-acceptor-donor organic phosphorescent emitters in symmetric and asymmetric architectures. Phenoxathiine (PX) is introduced as an alternative donor to thianthrene (TA), combined with benzophenone (BP) or pyridine (Py) as acceptors. Through photophysical characterization, quantum chemical simulations, and PLT device testing, we identify structure-property relationships and, in particular, investigate the impact of the individual moieties on the emission properties and stability. The RTP emission wavelength is primarily tunable through the donor moiety: PX-based emitters emit sky-blue ($\lambda_P \approx 480$ nm), while TA-based emitters emit in the green ($\lambda_P \approx 520$ nm) due to an increased Stokes shift. The acceptor unit strongly influences the phosphorescence quantum yield, with Py-based emitters systematically outperforming BP-based ones. All newly synthesized PX-containing emitters show sufficient performance in PLT devices, though with reduced photostability compared to TA-based counterparts. Together, these results demonstrate that systematic donor-acceptor design enables predictable control over RTP emission properties, advancing the rational development of high-performance RTP-based photonic devices.

## Introduction

Purely organic materials showing efficient and persistent emission via room temperature phosphorescence (RTP) allow the design of minimalistic yet powerful technological solutions for sensing, bioimaging, information storage, and safety applications using the photonic concept of digital luminescence. [1–10] The oxygen-sensitivity of RTP is used to locally switch between 'off' and 'on' states. With the design of programmable luminescent tags (PLTs), we recently reported a novel photonic device architecture that is well suited for various labelling and information exchange applications and even allows a biodegradable design.[11,12] As depicted in **Fig. 1a**, PLTs are three-layer devices based on substrate, oxygen blocking layer, and emissive layer with the latter consisting of sufficiently rigid polymer host and diluted purely organic phosphorescent emitter. Being fabricated from spin coating under ambient conditions, the emissive layer contains molecular oxygen that initially fully quenches the RTP emission. Upon excitation of the emitter molecules by illumination with UV light, controlled photoconsumption of the molecular oxygen is triggered leading to an activation of RTP emission when the oxygen concentration is sufficiently low (cf. **Fig. 1b**). With illumination being applied onto the entire PLT, these systems provide a promising solution for threshold UV dosimetry.[13] In contrast, by using masks or UV lasers as demonstrated in Figure 1c, activation can be triggered only locally leading to a photonic device for information storage.[11,14] By heating the activated PLT, the oxygen transmission of the blocking layer is increased and the information can be erased. More than 50 cycles of writing and erasing of information are feasible without significant losses of contrast.[11,15]

Despite these promising results, many challenges remain to exploit the full potential of PLTs and RTP-based photonic devices in general – most of them being connected to the development of specifically tailored efficient RTP emitters. Most promising RTP emitters, characterized by strong phosphorescence quantum yield $\varphi_{\mathrm{P}}$, low to negligible fluorescence quantum yield $\varphi_{\mathrm{F}}$ in the visible range, and sufficient photostability under UV irradiation, are regularly a byproduct of the search for emitters with efficient thermally-activated delayed fluorescence (TADF) and, thus, follow the donor-acceptor (D-A) design motif.[16–18] However, emitter development usually does not follow a systematic design approach, which complicates the derivation of structure-property relationships. Starting from the promising RTP emitter 4,4′-dithianthrene-1-yl-benzophenone (**BP-2TA**), we have recently systematically varied the electron-accepting character of the central acceptor moiety.[15] While it turned out that the electrophilicity does not systematically render the emission properties, we found that sufficient orbital hybridization in contrast with the charge transfer (CT) state character beneficial for TADF emission might lead to improved RTP emission as observed for 3,5-di(thianthren-1-yl)pyridine (**Py-2TA**, cf. **Fig. 1**). Despite the variations in chemical and electronic structures of the molecules studied, their emission occurs in a similar wavelength range with weak blue fluorescence ($\lambda_F \approx 450$ nm) and pronounced green RTP ($\lambda_P \approx 520$ nm), still limiting the exploitation of the full potential of PLTs and digital luminescence as a design motif.

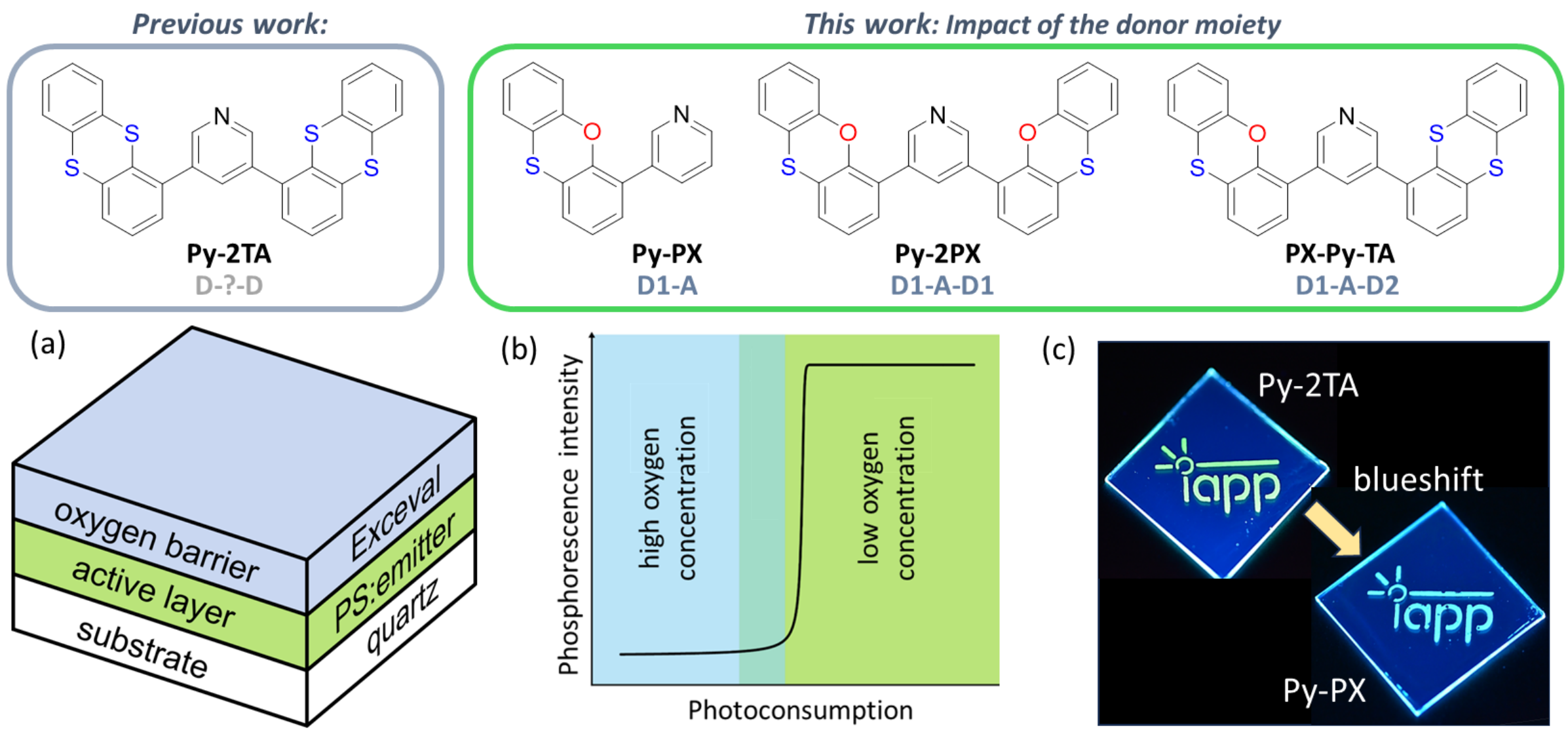

**Figure 1.** Systematic variation of the design space of RTP emitters following the donor-acceptor design motif with pyridine (**Py**) as acceptor unit as well as phenoxathiine (**PX**) and thianthrene (**TA**) as donor units. (a) Three-layer structure of PLTs with the materials used within this work. (b) Typical oxygen-sensitive activation of phosphorescent emission in PLTs with ongoing photoconsumption of molecular oxygen within the active layer upon excitation of the emitter molecules. (c) Comparison of the established emitter **Py-2TA** with the new emitter **Py-PX** illustrating the blueshift in emission of emitters compounds only containing **PX** as donor moiety.

Following this systematic evaluation of the design space of D-A emitter materials with tailored RTP emission and minimized fluorescence, we analyze the impact of the donor moiety in this contribution and investigate how the properties of the individual building blocks influence the photophysics of the composite emitter. So far, **Py-2TA** and **BP-2TA** with thianthrene (**TA**) as donor, and pyridine (**Py**) and benzophenone (**BP**) as acceptors have demonstrated the best performance as RTP emitters.[15] Accordingly, **Py** and **BP** are again used as core acceptor units and phenoxathiine (**PX**) is introduced as alternative donor (cf. **Fig. 1**). In contrast to **TA**[11,19–24], only a few studies exist on **PX** as a building block for RTP emitters.[25–28] **PX** is a heterocyclic compound containing sulphur and oxygen in the central core that reduces the heavy-atom effect on the RTP properties in comparison to **TA** (two sulphur in the central core) and exhibits an asymmetry in the conformation and electronic structure. A higher electronegativity of **PX** compared to **TA** has a significant impact on the electronic structure of composite molecules resulting into different emission behavior.[27] Recently, two new molecules following the D-A structural motif consisting of **PX** as a donor and **BP** as an acceptor were published and show sky-blue RTP, while asymmetric compounds that have phenothiazine or dimethylacridan donor units show TADF properties.[25] RTP molecules naming 4-phenoxathiin-4-yl-benzophenone (**PB**) and 4,4′-diphenoxathiin-4-yl-benzophenone (**DPB**) were independently discovered in our work, discussed here and named as **BP-PX** and **BP-2PX**.

Within this manuscript, we systematically explore the chemical space of donor-acceptor RTP emitters ranging from molecules with one (D-A) to two (D-A-D) donor moieties including symmetric and asymmetric designs. By combining experimental photophysical analysis with quantum chemical simulations, and application in PLTs, systematic insights in the interplay between chemical design, electronic structure, and photophysical properties are gained. Our results systematically demonstrate how the choice of donor and acceptor groups influences key properties of the composite emitter materials, thereby paving the way for the rational design of efficient RTP emitters.

## Results and discussion

### *Synthesis and crystal structures*

The synthesis of the target compounds was conducted *via* Suzuki-Miyaura cross-coupling reactions. The synthetic routes for compounds: *3-(phenoxathiin-4-yl)pyridine* (**Py-PX**), *(4-(phenoxathiin-4-yl)phenyl)(phenyl)methanone* (**BP-PX**), *3,5-di(phenoxathiin-4-yl)pyridine* (**Py-2PX**), *bis(4-(phenoxathiin-4-yl)phenyl)methanone* (**BP-2PX**), *3-(Phenoxathiin-4-yl)-5-(thianthren-1-yl)pyridine* (**PX-Py-TA**), and *(4-(Phenoxathiin-4-yl)phenyl)(4-(thianthren-1-yl)phenyl)methanone* (**PX-BP-TA**) are outlined in **Scheme 1**. Detailed descriptions are provided in the Supporting Information (SI) in section S1. The yields of the reactions for all compounds ranged between 40% and 90%, enabling large-scale synthesis. **BP-PX** and **BP-2PX** have been synthesized independently by the group of J.V. Grazulevicius following a similar yet slightly different synthetic route, resulting in a higher yield.[25] Asymmetrical compounds **PX-Py-TA** and **PX-BP-TA** were synthesized *via* two-step synthetic routes presenting an efficient way to obtain asymmetrical RTP materials. The chemical structures of all six compounds and intermediates were confirmed by $^{1}H$ and $^{13}C$ NMR analysis and mass spectrometry as presented in Section S2 in the Supporting Information (SI).

*Synthesis of D-A and D-A-D molecules*

**Py-PX**, 93 %  **BP-PX**, 50 %  **Py-2PX**, 49 %  **BP-2PX**, 40 %

*Synthesis of asymmetric molecules*

**Br-Py-TA**, 35 %  **PX-Py-TA**, 65 %

**Br-BP-TA**, 44 %  **PX-BP-TA**, 46 %

**Scheme 1.** Synthetic routes for **Py-PX**, **BP-PX**, **Py-2PX**, **BP-2PX**, **PX-Py-TA** and **PX-BP-TA** and corresponding yield after purification. Reaction conditions: (I) – 3 mol % of $Pd(PPh_3)_2$, $K_2CO_3$ aq.,THF, 75 °C; (II) – 5 mol % of $Pd(PPh_3)_2$, $K_2CO_3$ aq.,THF, 75 °C.

For compounds **Py-PX**, **BP-PX**, **Py-2PX** and **PX-BP-TA** single crystals were obtained by slow vapor diffusion into a saturated solution of the compound (see **Fig. 2**, S3). None of the compounds co-crystallized with solvent molecules, allowing for an extraction of the unperturbed crystal packing. In the case of **Py-PX** and **BP-PX**, the conformations of the molecules are comparably flat, indicated by more acute dihedral angles between donor and acceptor unit. In contrast, the dihedral angles in **Py-2PX** and **PX-BP-TA** are slightly larger, indicating higher strain, caused by increased steric bulk of two acceptor moieties. In our previous report[15] we highlighted that two acceptor molecules typically show one acceptor moiety pointing backwards and one forwards in the global energy minimum. In the case of **Py-2PX** this behavior is rediscovered. Interestingly, in **PX-BP-TA**, both acceptor moieties point into the same direction of the molecule, further demonstrating the high degree of flexibility and thus a low energy barrier for rotational motion.[15] This is also resembled by a rather high variation of dihedral angles for this compound.

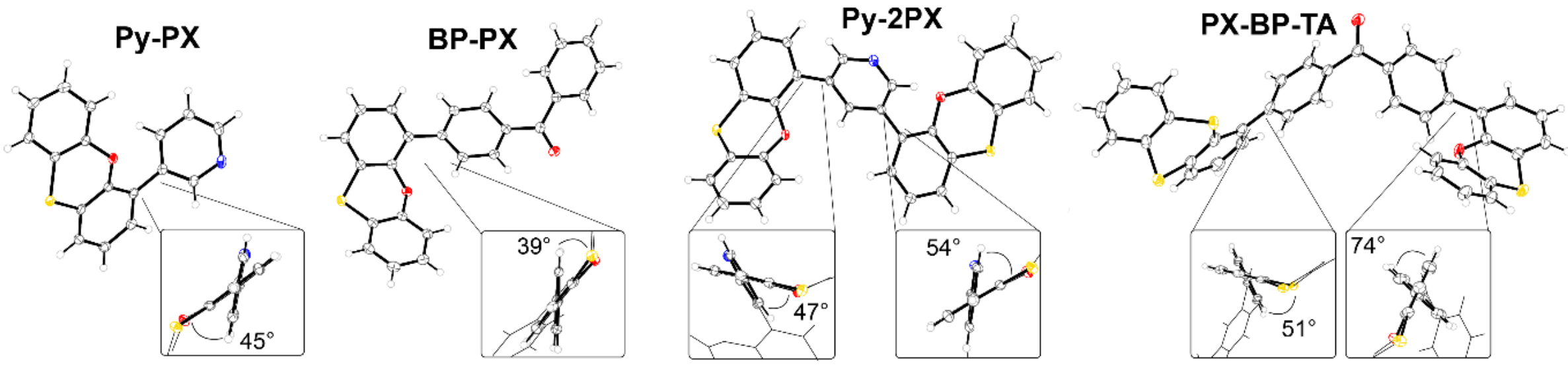

**Figure 2.** Molecular structures of **Py-PX**, **BP-PX**, **Py-2PX** and **PX-BP-TA**. Crystal structure information is provided in the SI in Section S3. Thermal ellipsoids are displayed at 50% probability.

*Thermal characteristics*

Organic emitter molecules regularly follow the D-A-D pattern, which is expected to provide improved thermal stability and higher glass transition temperature compared with their D-A analogs, which is especially relevant for application in vacuum-processed devices.[19] Results from simultaneous thermal analysis (STA) presented in Section S4 in the SI for the D-A and D-A-D compounds analyzed here are in agreement with this expectation. A comparison between **Py-2PX** and **Py-2TA** shows a relatively similar behavior with decomposition temperatures $T_d$ of 272 °C and 289 °C as well as melting points $T_m$ of 210 and 153 °C, respectively. The higher melting point of **Py-2PX** indicates stronger molecular interactions. Surprisingly, both asymmetric compounds **PX-Py-TA** and **PX-BP-TA** show the highest decomposition temperatures $T_d$ of 367 °C and 309 °C, respectively. In summary, the properties of all composite emitters appear to be sufficient for physical vapor deposition.

*Photophysics of building blocks: PX versus TA*

To evaluate the characteristics of the systematically varied RTP emitters in detail, we first discuss the photophysical properties of the building blocks used based on spectroscopic analysis and quantum chemical simulations. The absorption and emission properties of **BP**, **TA**, and **PX** diluted at 5 wt% in PS films are depicted in **Fig. 3** and summarized in Tab. 1. By comparing photoluminescence spectra under ambient and nitrogen atmosphere, pronounced RTP emission can be observed for **PX** ($\lambda_P$ = 480 nm, $\varphi_N$ = 16%) and **TA** ($\lambda_P$ = 520 nm, $\varphi_N$ = 28% vs. $\varphi_{air}$ = 4%), which is also confirmed by the delayed emission spectra.

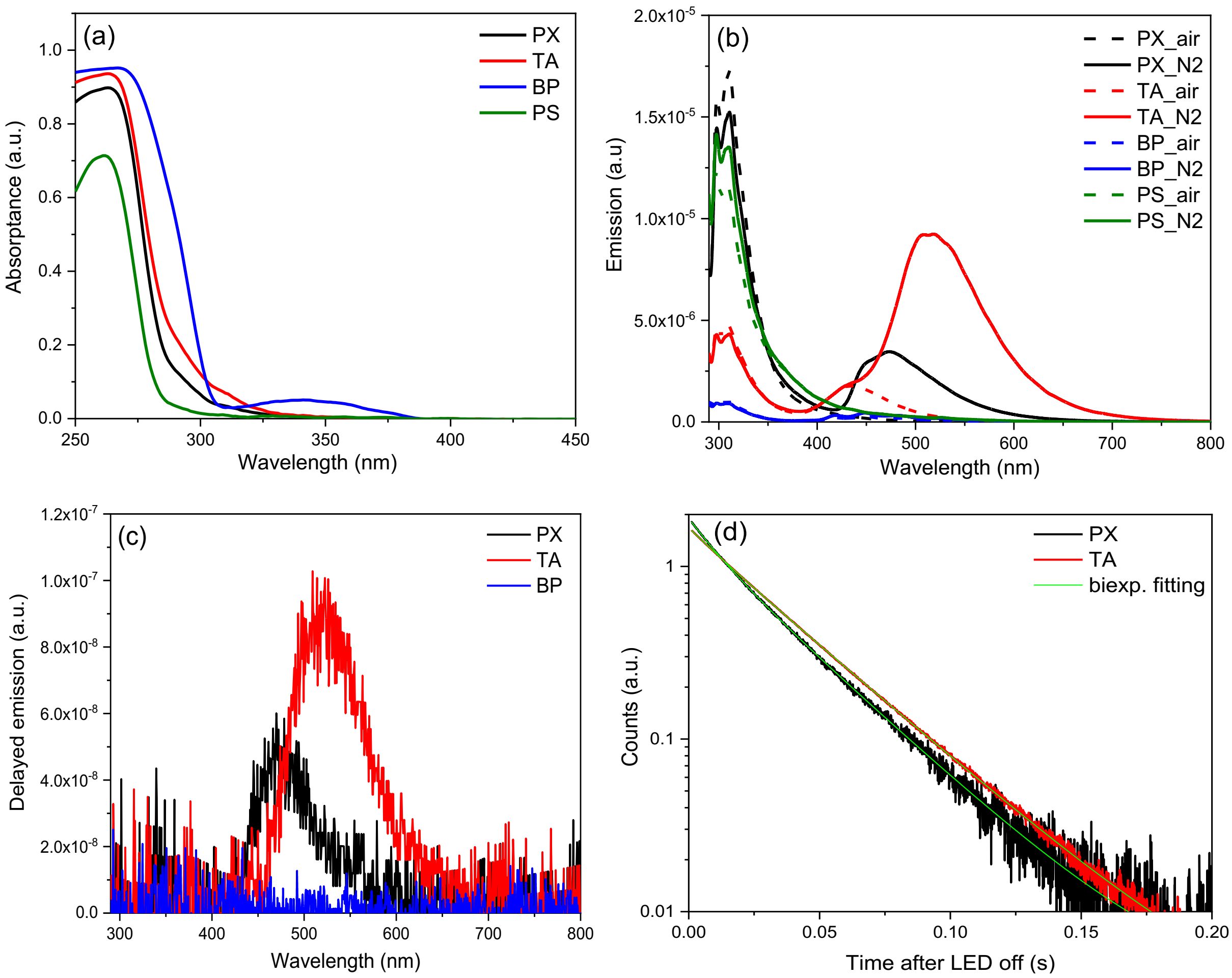

**Figure 3.** Photophysics of **PX**, **TA**, and **BP**. (a) Absorptance spectra in PS. (b) Emission spectra in PS with $\lambda_{\text{exc}} = 275$ nm under aerated (dashed lines) and nitrogen atmosphere at room temperature. (c) Corresponding delayed spectra with distinct emission signals only by **PX** and **TA**. (d) Phosphorescence decays of **PX** and **TA** in PS at room temperature under nitrogen atmosphere. Biexponential fit functions used to extract the phosphorescence lifetimes are presented as green lines. The strong emission band in the range from 290 nm to 350 nm corresponds to the fluorescent emission of PS.

Their phosphorescence lifetime is similar with 28 and 29 ms, respectively. While **TA** also demonstrates considerable fluorescence, the fluorescence from **PX** is so weak that it is not distinguishable from the fluorescence tail of the PS host. Emission spectra of the molecules in THF (cf. Fig. S24) show that not only the phosphorescence of **PX** is blue-shifted with respect to **TA** (200 meV), but also the fluorescence (500 meV). It needs to be noted that the high PLQY in nitrogen for **TA** is in contrast with previous studies in PMMA using 340 nm excitation and reporting values of approximately 6 %.[11] However, the low absorption of **TA** above 325 nm puts the accuracy of those results into question. With the detailed photophysical characterization of **TA** provided here, it shows similar performance to the best emitter **Py-2TA** and even outperforms **BP-2TA** with respect to PLQY.

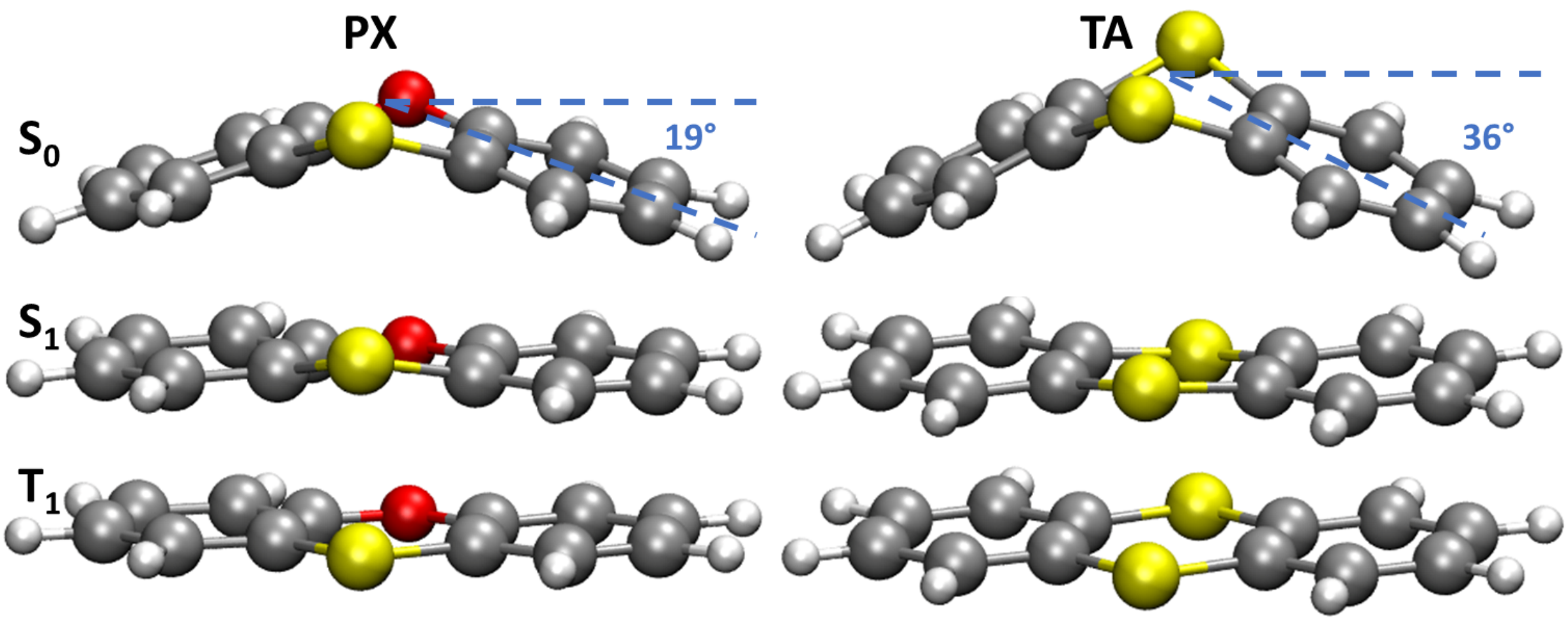

**Figure 4.** Relaxed molecules structures of **PX** and **TA** in their $S_0$, $S_1$, and $T_1$ states with bending angle in the relaxed $S_0$ state. Upon excitation, both molecules show widely planar structures. Colors of elements: C in silver, S in yellow, O in red.

The reason for the difference in emission wavelength between **PX** and **TA** can be understood from time-dependent density functional theory (TD-DFT) simulations. Based on the excited state analysis in the relaxed ground state geometry, both molecules appear to have energetically similar $S_0 \rightarrow S_1$ (4.60 eV for **PX**, 4.68 eV for **TA**) and $S_0 \rightarrow T_1$ (3.71 eV for **PX**, 3.80 eV for **TA**) transition energies. Mind that these computational values are not including polarization effects and therefore systematically shifted with respect to experimental values.[15,29] For both molecules, the $S_0 \rightarrow S_1$ transition is dark and the first transition with considerable oscillator strength ($\nu > 0.05$) is $S_0 \rightarrow S_6$ for **PX** and $S_0 \rightarrow S_4$ for **TA**. The bent relaxed $S_0$ geometry is planarized in both the relaxed $S_1$ and $T_1$ state as visualized in **Fig. 4** due to a change in the hybridization of the sulfur atom. Only the sulfur atoms are partially still slightly shifted out-of-plane. With the smaller bending angle for **PX** in the $S_0$ geometry (19° vs. 36° for **TA**), the relaxation energy is smaller as well, which results in a smaller Stokes shift. Comparison of transition energies in the different relaxed structures from TD-DFT gives a difference by approx. 300 meV for both fluorescence and phosphorescence, which is in agreement with the experimental observations – especially considering that the simulations do not explicitly consider effects of the polymer host on the molecular relaxations.

*Photophysics of newly synthesized materials*

**PX** and especially **TA** might represent efficient phosphorescent emitters themselves, but their absorption is relatively low and limited to deep UV below 300 nm (**Fig. 3a**). For application in PLTs, covering a broader absorption range is beneficial for applications, e.g. as UV sensors. In addition, their low molecular weight makes processing difficult using physical vapor deposition. Application of **TA** in D-A-D compounds has also resulted in reduced fluorescence. All assembled emitter materials show increased absorption between 275 and 350 nm with similar features independent from their specific molecular structure as depicted in **Fig. 5**, Fig. S26, and Fig. S27.

**Table 1.** Emission properties of the RTP emitters in PS at room temperature ($\lambda_{\mathrm{exc}} = 275$ nm).

| Compounds | $\lambda_{max}$ in air[a], nm | $\lambda_{max}$ in $N_2$[b], nm | $\tau_P$[c], ms | PLQY in air/$N_2$[d], % |
|---|---|---|---|---|
| **BP** | 425[e] | 425[e] | - | / |
| **TA** | 430 | 520 | 29 | 4/28 |
| **PX** | - | 480 | 28 | -[f]/16 |
| **Py-2TA**[15] | 440 | 530 | 28 | 2/25 |
| **Py-PX** | 390 | 480 | 29 | 1/21 |
| **Py-2PX** | 390 | 480 | 30 | 2/20 |
| **PX-Py-TA** | 430 | 520 | 27 | 2/21 |
| **BP-2TA**[15] | 415 | 520 | 24 | 1/16 |
| **BP-PX** | 430 | 480 | 88 | 1/9 |
| **BP-2PX** | 430 | 480 | 73 | 1/10 |
| **PX-BP-TA** | 430 | 520 | 38 | 1/8 |

[a] Wavelength of the emission maximum under aerated atmosphere. [b] Wavelength of the emission maximum under $N_2$ atmosphere. [c] Phosphorescence lifetime extracted from delayed spectroscopy (cf. Figs 7d & S29b) via a triexponential fit. [d] PLQY under aerated and $N_2$ atmosphere. [e] The wavelength does not refer to the emission maximum, but to the 0-0 emission peak as the emission spectrum of BP shows a pronounced vibronic progression (cf. Fig. S25). [f] No distinct fluorescent emission feature by PX is observed (cf. Fig. 3b).

The emission spectra of **Py-2TA**, **Py-PX**, **Py-2PX**, and **PX-Py-TA** diluted in PS under ambient and nitrogen atmosphere at room temperature are presented in **Fig. 6** and for **BP-PX**, **BP-2PX**, **PX-BP-TA** and **BP-2TA** in Fig. S28 for $\lambda_{\mathrm{exc}} = 275$ nm. The emission properties of all materials are summarized in Tab. 1. In comparison to **Py-2TA** exhibiting green RTP, D–A (**Py-PX** and **BP-PX**) and D–A–D (**Py-2PX** and **BP-2PX**) compounds with symmetrical architectures exhibit pronounced sky-blue RTP emission with a maximum wavelength at 480 nm, consistent with the RTP emission of **PX** alone. No significant differences between D-A and corresponding D-A-D emitter materials are observed. The spectral position and shape of the composite emitters shows strong similarity to the individual donor moieties indicating a $^3$LE phosphorescent emission in agreement with previous studies.[11] D-A-D compounds with asymmetrical architectures **PX-Py-TA** and **PX-BP-TA** show green RTP emission with a maximum wavelength at 520 nm consistent with the characteristics of **TA** and its lower triplet state compared with **PX**. In contrast, it appears that the PLQY of the asymmetric emitters is dictated by **PX** as it potentially introduces further loss mechanisms or reduces the ISC rate. In summary, the emission wavelength of the composite emitter materials is dictated by the donor moiety used potentially allowing for a stronger variation of the emission wavelengths by using suitable donor moieties.

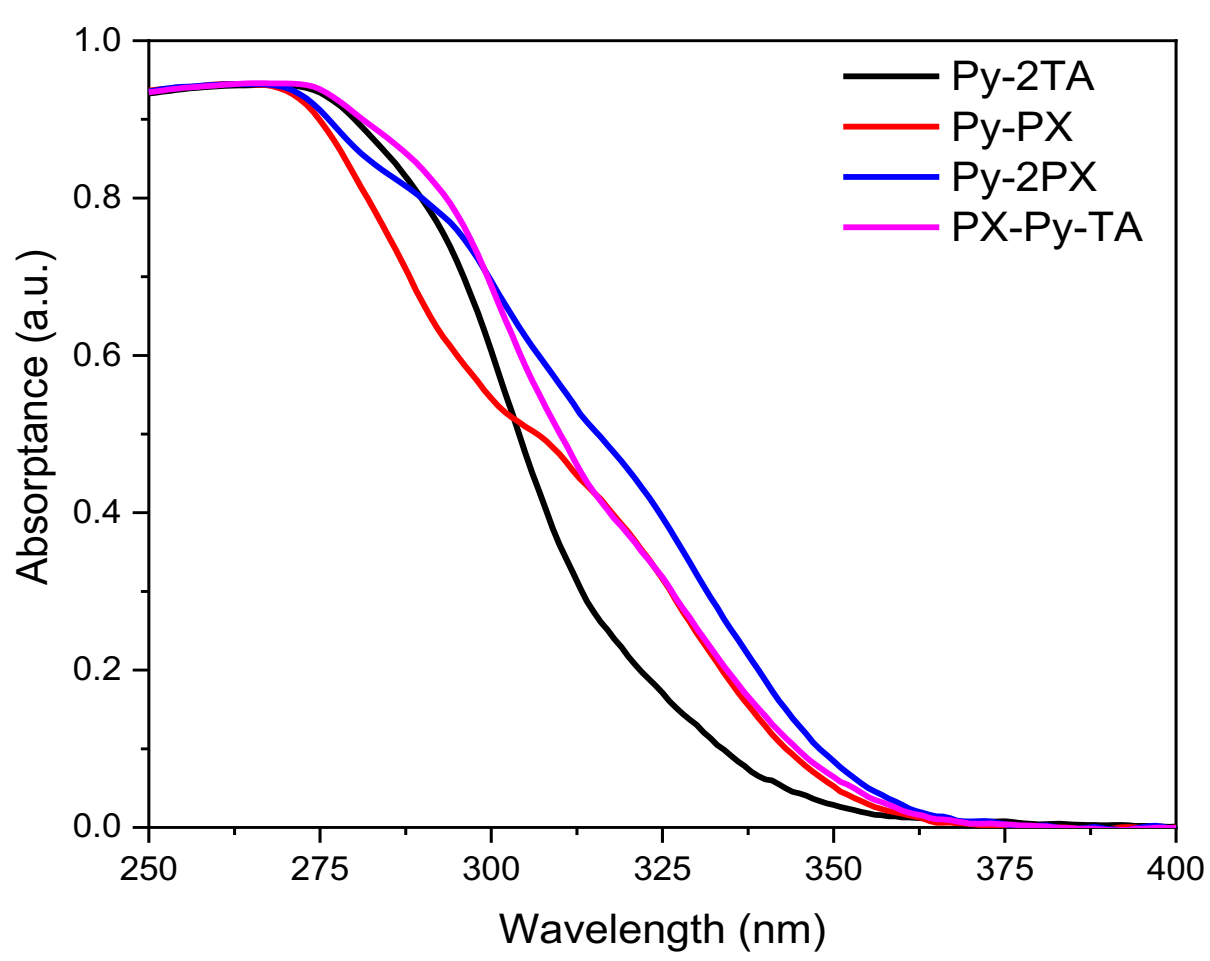

**Figure 5.** Absorptance spectra of **Py-2TA**, **Py-PX**, **Py-2PX**, and **PX-Py-TA** in PS.

The choice of the acceptor unit is significantly affecting the PLQY of phosphorescence and fluorescence. **Py** is only slightly reducing the phosphorescent PLQY of **TA** and even increasing it for **PX** combined with a systematic reduction of fluorescence based on a pronounced CT state character of the $S_1$ state. In contrast, the connection with **BP** causes a systematic reduction of the PLQY values for both fluorescence and phosphorescence. We assume that the flexible nature of **BP** introduces more non-radiative losses compared to the relatively compact and stiff nature of **Py**, which can even be able to promote the phosphorescent emission of the donor moieties. However, the surprisingly long phosphorescent lifetimes of **BP-PX** and **BP-2PX** contradict this assumption. Here, future studies are required to gain a deeper understanding of the impact of the donor and acceptor units on composite RTP emitters.

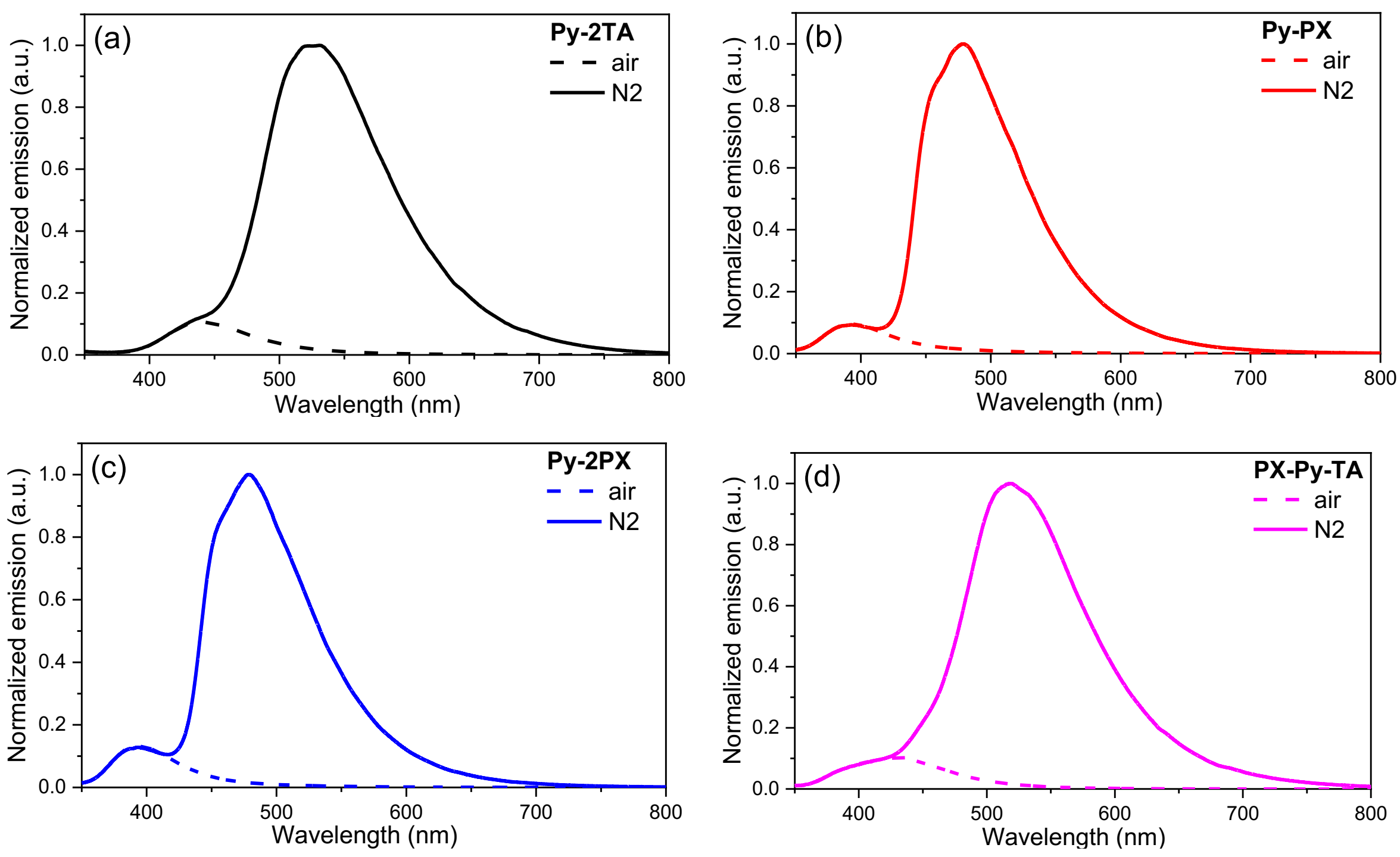

**Figure 6.** Emission spectra ($\lambda_{exc}$ = 275 nm) of **Py-2TA** (a), **Py-PX** (b), **Py-2PX** (c), and **PX-Py-TA** (d) in PS at room temperature under aerated (dashed lines) and nitrogen atmosphere.

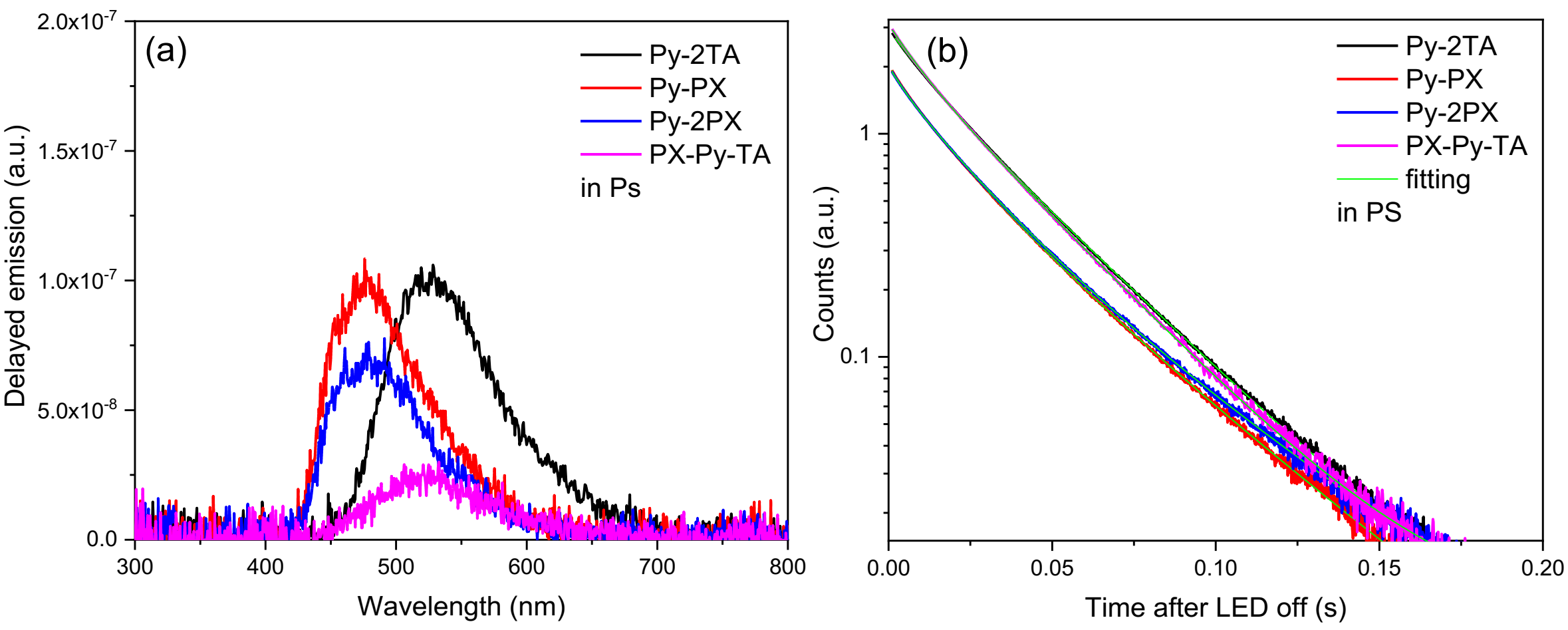

**Figure 7.** Persistent luminescence of the emitters: delayed spectra (a) and corresponding phosphorescence decay (b) of **Py-2TA**, **Py-PX**, **Py-2PX**, and **PX-Py-TA** in PS at room temperature under nitrogen atmosphere collected at a delay time of 10 ms, showing only the phosphorescence ($\lambda_{exc}$ = 275 nm). Biexponential fit function (**Py-2TA** and **PX-Py-TA**) and triexponential fit functions (**Py-PX** and **Py-2PX**) and used to extract the phosphorescence lifetimes are presented as green lines in (b).

*Application in Programmable Luminescent Tags*

All compounds incorporating **PX** as donor unit show sufficient photophysical characteristics for application in PLTs. For single use, e.g. for UV dosimetry[13], their first activation is most critical. The performance for all **Py**-based emitters as well as **Py-2TA** as reference and **TA** and **PX** is presented in **Fig. 8a**. All emitters demonstrate fully functional devices as can be seen as well in Fig. S30 in the SI. The PLTs cover a relatively large range of activation doses $D_{\text{act}}$ from 54.9 mJ/cm² for **Py-PX** to 103.4 mJ/cm² for **PX** illustrating the tunability of PLTs by choosing different emitter materials (Table 2). Mind that the activation dose can be further modified based on the emitter concentration[13] and host material[15]. The best brightness contrast between off and on state is obtained for **Py-2TA**, while the lowest contrast is obtained for the individual donor molecules **TA** and **PX**, for which the low absorption is only partially compensated by a higher total amount of molecules in the active layer. Still, for all emitters the contrast should be sufficient for efficient read-out especially due to their weak fluorescence.

Reusability of PLTs is tested over ten cycles of writing, *i.e.* activation through UV excitation, and erasing, *i.e.* heating the PLT on a hot plate to allow oxygen to pass the oxygen barrier (see Section S1 in the SI for details). The maximum brightness after each activation is depicted in **Fig. 8b**. While **TA** and **TA**-based emitters show only a weak decrease in brightness by approx. 8%, **PX** and **PX**-based emitters are more prone to degrade with a decrease of over 50%. The performance of the asymmetric emitter **PX-Py-TA** is in between with a decrease by almost 20%. Currently, it can only be speculated if **PX** is more likely to perform oxidation reactions with the singlet oxygen formed or if it is more sensitive to UV illumination especially considering the higher energy of its excited triplet state. Overall, the results suggest that the choice of donor group is also strongly influencing the stability of composite RTP emitters.

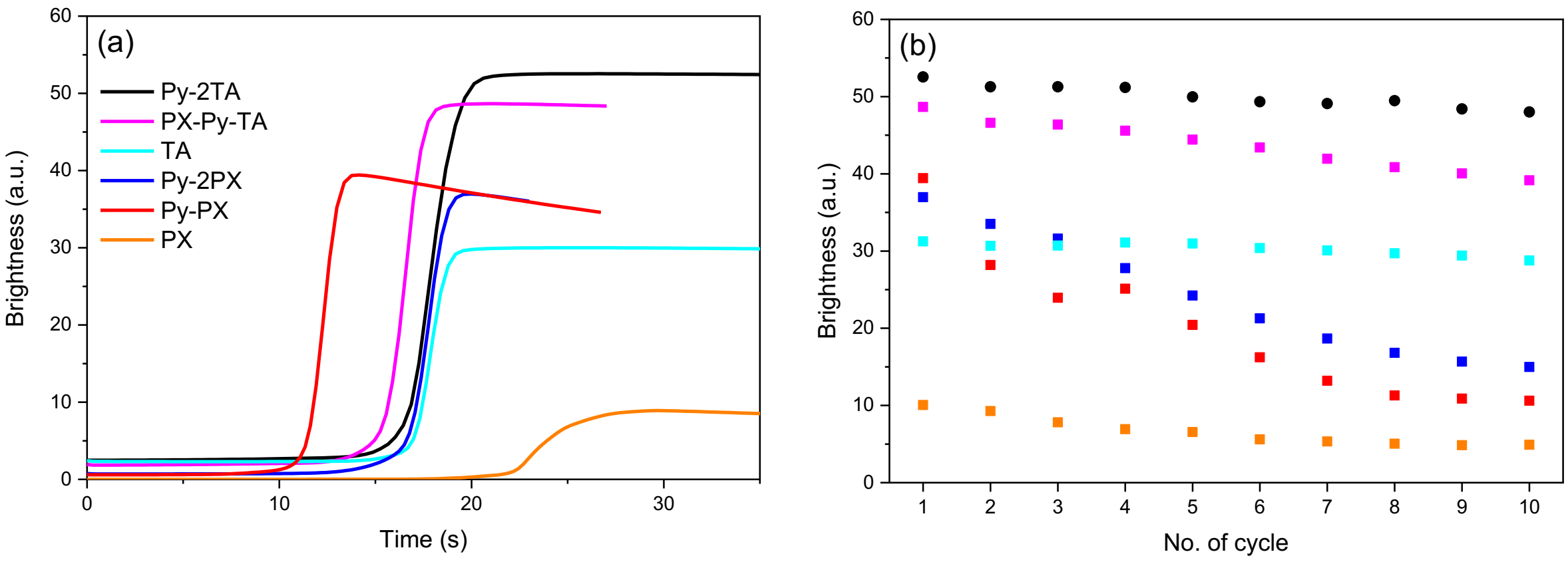

**Figure 8.** Performance of PLTs using **Py-2TA**, **Py-PX**, **Py-2PX**, and **PX-Py-TA** as RTP emitters. (a) Photoluminescence response over UV-illumination time for the first activation cycle. (b) Maximum photoluminescence intensity over ten cycles of writing and erasing. The brightness is defined as the averaged signal of all three color channels from the RGB camera. In all cases the values are averaged over three PLTs, and 420 x 420 pixels à 42.3 x 42.3 µm² ($\lambda_{\text{exc}}$ = 280 nm).

**Table 2.** Key performance metrics of different RTP emitters in PLTs at 5 wt% in PS.

| Emitter | Py-2TA | PX-Py-TA | TA | Py-2PX | Py-PX | PX |
|---|---|---|---|---|---|---|
| $t_{act}$[a], s | 18.4 | 16.5 | 17.8 | 17.6 | 12.2 | 23.1 |
| $D_{act}$[b], mJ/cm$^2$ | 82.8 | 74.3 | 80.1 | 79.2 | 54.9 | 103.4 |
| $s_{10}$[c], % | 91.4 | 80.5 | 92.0 | 40.5 | 26.9 | 48.8 |

[a] The activation time $t_{act}$ of a PLT is defined by the time required to reach half of its maximum phosphorescence brightness under constant UV-illumination (cf. Fig. 8a). [b] The activation dose ($D_{act}$) is calculated as $D_{act} = I_0 \cdot t_{act}$, where $I_0 = 4.5$ mW/cm². [c] The brightness retention at 10th cycle ($s_{10}$) is defined as the relative maximum brightness after 10 cycles of activation and deactivation with respect to the maximum brightness after the first activation.

**Conclusions**

Although many RTP emitters follow a donor-acceptor or donor-acceptor-donor structural motif, an incomplete understanding of the structure-property relationships regularly hinders the application-specific tailoring of the emitter's properties. By systematically analyzing the design space of emitters using **PX** and **TA** as donor moieties and **BP** and **Py** as acceptor groups, essential information on promoting the rational design of RTP emitters is obtained.

The emission wavelength of newly developed emitters is directly tunable through the donor moiety. **PX**-based compounds emit sky-blue phosphorescence (480 nm), while **TA**-based compounds emit in the green (520 nm), consistent with the localized triplet excited state character ($^3$LE) of individual donor moieties **PX** and **TA**. Quantum chemical simulations attribute the blue shift of **PX** relative to **TA** to the former's smaller bending angle in the $S_0$ geometry (19° vs. 36°), which reduces the energy loss during geometric relaxation following excitation and results in a smaller Stokes shift. In asymmetric D-A-D structures combining both donors, the emission wavelength is determined by **TA** as the lower-energy triplet state donor, while the quantum yield is compromised by the presence of **PX**. These results establish the donor moiety as the key design parameter for emission wavelength tuning, suggesting that introduction of further donor units could extend the accessible emission range beyond the blue-green region.

The choice of acceptor unit strongly affects the phosphorescence quantum yield. **Py**-based emitters systematically preserve or even enhance the phosphorescent efficiency of the donor units. **BP**-based emitters, by contrast, show reduced phosphorescence quantum yields, likely due to the higher conformational freedom of **BP** introducing additional non-radiative decay pathways. In all cases, fluorescence is sufficiently suppressed through a pronounced charge-transfer character of the $S_1$ state.

All **PX**-containing emitters were successfully implemented in PLT devices with tunable UV activation doses, demonstrating the flexibility of PLT design through emitter selection. However, the photostability of **PX**-based emitters is significantly reduced over repeated write-erase cycles, with brightness losses significantly exceeding those of their **TA**-based counterparts, indicating a higher susceptibility of **PX** to photochemical degradation or oxidation reactions.

These findings collectively demonstrate that the rational combination of donor and acceptor building blocks offers a powerful and systematic route toward tuning the emission wavelength, quantum yield, and device stability of organic RTP emitters, providing a valuable design framework for the development of next-generation PLTs and further RTP-based photonic applications.

**Acknowledgements**

Funded by the European Union (ERC, SLOWTONICS, 101089234). Views and opinions expressed are however those of the author(s) only and do not necessarily reflect those of the European Union or the European Research Council Executive Agency. Neither the European Union nor the granting authority can be held responsible for them. In addition, Uliana Tsiko thanks the Special Research Fellowship of the Alexander von Humboldt Foundation for financial support in frame of the project “ArtOfRTP”. K.S.S. thanks the Center for Information Services and High Performance Computing (ZIH) at TUD Dresden University of Technology for the use of computational facilities.

## References

(1) Gao, M.; Wu, R.; Zhang, Y.; Meng, Y.; Fang, M.; Yang, J.; Li, Z. New Molecular Photoswitch Based on the Conformational Transition of Phenothiazine Derivatives and Corresponding Triplet Emission Properties. *J. Am. Chem. Soc.* **2025**, *147* (3), 2653–2663. https://doi.org/10.1021/jacs.4c14920.

(2) Zhang, L.; Li, J.; Zhang, Y.; Dai, W.; Zhang, Y.; Gao, X.; Liu, M.; Wu, H.; Huang, X.; Lei, Y.; Ding, D. White Light-Excited Organic Room-Temperature Phosphorescence for Improved in Vivo Bioimaging. *Nat. Commun.* **2025**, *16* (1), 1–13. https://doi.org/10.1038/s41467-025-59367-0.

(3) Rui Du; Zhengshuo Wang; Zhipeng Zhao; Huilong Liu; Shouchang Jiao; Yi Wu; Wenhui Li; Hua Yuan; Hanlin Ou; Dan Ding. Multicolor 3D Afterglow Structures with High Precision and Ultralong Lifetimes Based on Carbazole-Doped Photocurable Resins. *Mater. Chem. Front.* **2025**, 1–11. https://doi.org/10.1039/d5qm00369e.

(4) Deng, Y.; Li, P.; Li, J.; Sun, D.; Li, H. Color-Tunable Aqueous Room-Temperature Phosphorescence Supramolecular Assembly. *ACS Appl. Mater. Interfaces* **2021**, *13* (12), 14407–14416. https://doi.org/10.1021/acsami.1c01174.

(5) Ma, F.; Wu, B.; Zhang, S.; Jiang, J.; Shi, J.; Ding, Z.; Zhang, Y.; Tan, H.; Alam, P.; Lam, J. W. Y.; Xiong, Y.; Li, Z.; Tang, B. Z.; Zhao, Z. Lone Pairs-Mediated Multiple Through-Space Interactions for Efficient Room-Temperature Phosphorescence. *J. Am. Chem. Soc.* **2025**, *147* (12), 10803–10814. https://doi.org/10.1021/jacs.5c02567.

(6) Zhao, Y.; Yang, J.; Liang, C.; Wang, Z.; Zhang, Y.; Li, G.; Qu, J.; Wang, X.; Zhang, Y.; Sun, P.; Shi, J.; Tong, B.; Xie, H. Y.; Cai, Z.; Dong, Y. Fused-Ring Pyrrole-Based Near-Infrared Emissive Organic RTP Material for Persistent Afterglow Bioimaging. *Angew. Chem. Int. Ed.* **2024**, *63* (5), e202317431. https://doi.org/10.1002/anie.202317431.

(7) Chen, Z.; Chen, X.; Ma, D.; Mao, Z.; Zhao, J.; Chi, Z. Synergetic Conformational Regulations in Ground and Excited States for Realizing Stimulus-Responsive and Wide-Tuning Room-Temperature Phosphorescence. *J. Am. Chem. Soc.* **2023**, *145* (30), 16748–16759. https://doi.org/10.1021/jacs.3c04725.

(8) Xiao, Y.; Li, J.; Song, Z.; Liao, J.; Shen, M.; Yu, T.; Huang, W. 3D Printable Materials with Visible Light Triggered Photochromism and Room Temperature Phosphorescence. *J. Am. Chem. Soc.* **2025**, *147* (24), 20372–20380. https://doi.org/10.1021/jacs.5c00976.

(9) Xiong, S.; Xiong, Y.; Wang, D.; Pan, Y.; Chen, K.; Zhao, Z.; Wang, D.; Tang, B. Z. Achieving Tunable Organic Afterglow and UV-Irradiation-Responsive Ultralong Room-Temperature Phosphorescence from Pyridine-Substituted Triphenylamine Derivatives. *Adv. Mat.* **2023**, *35* (28), 2301874. https://doi.org/10.1002/adma.202301874.

(10) Lei, Y.; Dai, W.; Li, G.; Zhang, Y.; Huang, X.; Cai, Z.; Dong, Y. Stimulus-Responsive Organic Phosphorescence Materials Based on Small Molecular Host-Guest Doped

Systems. *J. Phys. Chem. Lett.* **2023**, *14* (7), 1794–1807. https://doi.org/10.1021/acs.jpclett.2c03914.

(11) Gmelch, M.; Achenbach, T.; Tomkeviciene, A.; Reineke, S.; Gmelch, M.; Achenbach, T.; Tomkeviciene, A.; Reineke, S. High-Speed and Continuous-Wave Programmable Luminescent Tags Based on Exclusive Room Temperature Phosphorescence (RTP). *Adv. Sci.* **2021**, *8* (23), 2102104. https://doi.org/10.1002/advs.202102104.

(12) Thomas, H.; Achenbach, T.; Hodgkinson, I. M.; Spoerer, Y.; Kuehnert, I.; Dornack, C.; Schellhammer, K. S.; Reineke, S. Room Temperature Phosphorescence from Natural, Organic Emitters and Their Application in Industrially Compostable Programmable Luminescent Tags. *Adv. Mat.* **2024**, *36* (26), 2310674. https://doi.org/10.1002/adma.202310674.

(13) Achenbach, T.; Will, P.-A.; Schellhammer, K. S.; Reineke, S. Tunable-Threshold UV Dosimetry with Programmable Luminescent Tags via Oxygen-Mediated Room-Temperature Phosphorescence. *Adv. Mater. Technol.* **2026**.

(14) Kantelberg, R.; Achenbach, T.; Kirch, A.; Reineke, S. In-Plane Oxygen Diffusion Measurements in Polymer Films Using Time-Resolved Imaging of Programmable Luminescent Tags. *Sci. Rep.* **2024**, *14* (1), 1–12. https://doi.org/10.1038/s41598-024-56237-5.

(15) Tsiko, U.; Fidelius, J.; Kaiser, S.; Thomas, H.; Bui Thi, Y.; Weigand, J. J.; Grazulevicius, J. V.; Schellhammer, K. S.; Reineke, S. Systematic Variation of the Acceptor Electrophilicity in Donor-Acceptor-Donor Emitters Exhibiting Efficient Room Temperature Phosphorescence Suited for Digital Luminescence. *Commun. Chem.* **2025**, *8* (1), 1–10. https://doi.org/10.1038/s42004-025-01620-0.

(16) Chen, C.; Huang, R.; Batsanov, A. S.; Pander, P.; Hsu, Y. T.; Chi, Z.; Dias, F. B.; Bryce, M. R. Intramolecular Charge Transfer Controls Switching Between Room Temperature Phosphorescence and Thermally Activated Delayed Fluorescence. *Angew. Chem. Int. Ed.* **2018**, *57* (50), 16407–16411. https://doi.org/10.1002/anie.201809945.

(17) Paredis, S.; Cardeynaels, T.; Deckers, J.; Danos, A.; Vanderzande, D.; Monkman, A. P.; Champagne, B.; Maes, W. Bridge Control of Photophysical Properties in Benzothiazole-Phenoxazine Emitters – from Thermally Activated Delayed Fluorescence to Room Temperature Phosphorescence. *J. Mater. Chem. C* **2022**, *10* (12), 4775–4784. https://doi.org/10.1039/d1tc04885f.

(18) Song, T.; Liu, H.; Ren, J.; Wang, Z. Achieving TADF and RTP with Stimulus-Responsiveness and Tunability from Phenothiazine-Based Donor−Acceptor Molecules. *Adv. Opt. Mater.* **2024**, *12* (1), 2301215. https://doi.org/10.1002/adom.202301215.

(19) Tomkeviciene, A.; Dabulienė, A.; Matulaitis, T.; Guzauskas, M.; Andruleviciene, V.; Grazulevicius, J. V.; Yamanaka, Y.; Yano, Y.; Ono, T. Bipolar Thianthrene Derivatives Exhibiting Room Temperature Phosphorescence for Oxygen Sensing. *Dyes and Pigments* **2019**, *170*, 107605. https://doi.org/10.1016/j.dyepig.2019.107605.

(20) Qiu, W.; Cai, X.; Chen, Z.; Wei, X.; Li, M.; Gu, Q.; Peng, X.; Xie, W.; Jiao, Y.; Gan, Y.; Liu, W.; Su, S. J. A “Flexible” Purely Organic Molecule Exhibiting Strong Spin-Orbital Coupling: Toward Nondoped Room-Temperature Phosphorescence OLEDs. *J. Phys. Chem. Lett.* **2022**, *13* (22), 4971–4980. https://doi.org/10.1021/acs.jpclett.2c01205.

(21) Pander, P.; Swist, A.; Turczyn, R.; Pouget, S.; Djurado, D.; Lazauskas, A.; Pashazadeh, R.; Grazulevicius, J. V.; Motyka, R.; Klimash, A.; Skabara, P. J.; Data, P.; Soloducho, J.; Dias, F. B. Observation of Dual Room Temperature Fluorescence-Phosphorescence in Air, in the Crystal Form of a Thianthrene Derivative. *J. of Phys. Chem. C* **2018**, *122* (43), 24958–24966. https://doi.org/10.1021/acs.jpcc.8b08329.

(22) Yang, Z.; Zhao, S.; Zhang, X.; Liu, M.; Liu, H.; Yang, B. Efficient Room-Temperature Phosphorescence from Discrete Molecules Based on Thianthrene Derivatives for Oxygen Sensing and Detection. *Front. Chem.* **2022**, *9*. https://doi.org/10.3389/fchem.2021.810304.

(23) Meng, Y.; Liu, W.; Liu, Z.; Gao, M.; Fang, M.; Yang, J.; Ma, D.; Li, Z. Pure Room Temperature Phosphorescence Emission in Nondoped OLEDs: Adjustable Oxidation States and Excited-State Modulation. *ACS Appl. Mater. Interfaces* **2024**, *16* (44), 60658–60665. https://doi.org/10.1021/acsami.4c13336.

(24) Chen, J.; Tian, H.; Yang, Z.; Zhao, J.; Yang, Z.; Zhang, Y.; Aldred, M. P.; Chi, Z.; Chen, J.; Zhao, J.; Tian, H.; Yang, Z.; Zhang, Y.; Chi, Z.; Aldred, M. P. A Multi-Stimuli-Responsive Molecule with Responses to Light, Oxygen, and Mechanical Stress through Flexible Tuning of Triplet Excitons. *Adv. Opt. Mater.* **2021**, *9* (2), 2001550. https://doi.org/10.1002/adom.202001550.

(25) Dabuliene, A.; Quignon, M.; Bezvikonnyi, O.; Keruckiene, R.; Abdella, M.; Simokaitiene, J.; Volyniuk, D.; Grazulevicius, J. V. Management of Triplet Electronic Excitations in Derivatives of Phenoxathiin and Benzophenone. *ACS Appl. Electron. Mater.* **2025**, *7*, 6137–6148. https://doi.org/10.1021/acsaelm.5c00863.

(26) Dvylys, L.; Keruckiene, R.; Volyniuk, L.; Volyniuk, D.; Grazulevicius, J. V. Emissive Tags Based on Three New Electron Acceptors and Phenoxathiin Containing Compounds with Enhancement of Room-Temperature Phosphorescence. *Chem. Engineer. J.* **2025**, *505*, 159362. https://doi.org/10.1016/j.cej.2025.159362.

(27) Pan, G.; Yang, Z.; Liu, H.; Wen, Y.; Zhang, X.; Shen, Y.; Zhou, C.; Zhang, S. T.; Yang, B. Folding-Induced Spin-Orbit Coupling Enhancement for Efficient Pure Organic Room-Temperature Phosphorescence. *J. Phys. Chem. Lett.* **2022**, *13* (6), 1563–1570. https://doi.org/10.1021/acs.jpclett.1c04180.

(28) Ma, H.; Fu, L.; Yao, X.; Jiang, X.; Lv, K.; Ma, Q.; Shi, H.; An, Z.; Huang, W. Boosting Organic Phosphorescence in Adaptive Host-Guest Materials by Hyperconjugation. *Nat. Commun.* **2024**, *15* (1), 1–7. https://doi.org/10.1038/s41467-024-47992-0.

(29) Schellhammer, K. S.; Li, T. Y.; Zeika, O.; Körner, C.; Leo, K.; Ortmann, F.; Cuniberti, G. Tuning Near-Infrared Absorbing Donor Materials: A Study of Electronic, Optical, and Charge-Transport Properties of Aza-BODIPYs. *Chemistry of Materials* **2017**, *29* (13), 5525–5536. https://doi.org/10.1021/acs.chemmater.7b00653.

(30) Oxford Diffraction / Agilent Technologies UK Ltd, CrysAlisPRO.

(31) Dolomanov, O. V; Bourhis, L. J.; Gildea, R. J.; Howard, J. A. K.; Puschmann, H. OLEX2: A Complete Structure Solution, Refinement and Analysis Program. *J. Appl. Cryst* **2009**, *42*, 339–341. https://doi.org/10.1107/S0021889808042726.

(32) Sheldrick, G. M. Foundations and Advances SHELXT-Integrated Space-Group and Crystal-Structure Determination. *Acta Cryst* **2015**, *71*, 3–8. https://doi.org/10.1107/s2053273314026370.

(33) Frisch, M. J.; Trucks, G. W.; Schlegel, H. B.; Scuseria, G. E.; Robb, M. A.; Cheeseman, J. R. Gaussian 16, Rev. C.01. *Gaussian Inc., Wallingford CT* **2016**.

(34) Neese, F. Software Update: The ORCA Program System — Version 6.0. *Wiley Interdiscip. Rev. Comput. Mol. Sci.* **2025**, *15* (2), e70019. https://doi.org/0.1002/wcms.70019.

(35) Yanai, T.; Tew, D. P.; Handy, N. C. A New Hybrid Exchange–Correlation Functional Using the Coulomb-Attenuating Method (CAM-B3LYP). *Chem. Phys. Lett.* **2004**, *393* (1–3), 51–57. https://doi.org/10.1016/j.cplett.2004.06.011.

(36) McLean, A. D.; Chandler, G. S. Contracted Gaussian Basis Sets for Molecular Calculations. I. Second Row Atoms, Z=11–18. *J. Chem. Phys.* **1980**, *72* (10), 5639–5648. https://doi.org/10.1063/1.438980.

(37) Krishnan, R.; Binkley, J. S.; Seeger, R.; Pople, J. A. Self-consistent Molecular Orbital Methods. XX. A Basis Set for Correlated Wave Functions. *J. Chem. Phys.* **1980**, *72* (1), 650–654. https://doi.org/10.1063/1.438955.

(38) Matsidik, R.; Komber, H.; Brinkmann, M.; Schellhammer, K. S.; Ortmann, F.; Sommer, M. Evolution of Length-Dependent Properties of Discrete n-Type Oligomers Prepared via Scalable Direct Arylation. *J. Am. Chem. Soc.* **2023**, *145*, 8444. https://doi.org/10.1021/jacs.3c00058

(39) Fidelius, J.; Schwedtmann, K.; Schellhammer, S.; Haberstroh, J.; Schulz, S.; Huang, R.; Klotzsche, M. C.; Bauzá, A.; Frontera, A.; Reineke, S.; Weigand, J. J. Convenient Access to π-Conjugated 1,3-Azaphospholes from Alkynes via [3 + 2]-Cycloaddition and Reductive Aromatization. *Chem* **2024**, *10* (2), 644–659. https://doi.org/10.1016/j.chempr.2023.10.016.

(40) Fischer, A.; Kaschura, F. *SweepMe! A multi-tool measurement software.* www.sweep-me.net .

(41) Salas Redondo, C.; Kleine, P.; Roszeitis, K.; Achenbach, T.; Kroll, M.; Thomschke, M.; Reineke, S. Interplay of Fluorescence and Phosphorescence in Organic Biluminescent

Emitters. *J. of Phys. Chem. C* **2017**, *121* (27), 14946–14953. https://doi.org/10.1021/acs.jpcc.7b04529

(42) Lakowicz, J. R. Principles of Fluorescence Spectroscopy. *Principles of Fluorescence Spectroscopy* **2006**, 1–954. https://doi.org/10.1007/978-0-387-46312-4

(43) De Mello, J. C.; Wittmann, H. F.; Friend, R. H. An Improved Experimental Determination of External Photoluminescence Quantum Efficiency. *Adv. Mat.* **1997**, *9* (3), 230–232. https://doi.org/10.1002/adma.19970090308.

(44) Fries, F.; Reineke, S. Statistical Treatment of Photoluminescence Quantum Yield Measurements. *Scie. Rep. 2019* **2019**, *9* (1), 1–6. https://doi.org/10.1038/s41598-019-51718-4.

**Supplementary Information**

**Table of Contents**

## S1. Materials and Methods

*Materials*

3-Bromopyridine (*Fluorochem*), 3,5-dibromopyridine (*Sigma-Aldrich*), 4-brombenzophenon (*Thermo Scientific*), 4,4′-dibrombenzophenon (*Alfa Aesar*), phenoxathiin-4-boronic acid (*Abcr*), thianthrene-1-boronic acid (*Sigma-Aldrich*), tetrakis(triphenylphosphine)palladium(0) (*Thermo Scientific*) and potassium carbonate (*Fisher Scientific*) were used for synthesis of the designed materials. All dry solvents were purchased commercially. Polystyrene ($M_W$ = 35,000 g/mol) obtained from Sigma-Aldrich and Exceval from Kuraray Europe GmbH were used for film fabrication.

*NMR analysis*

NMR spectra were measured on a Bruker *AVANCE III HDX, 500 MHz Ascend* ($^{1}$H (500.13 MHz), $^{13}$C (125.75 MHz)). All $^{13}$C NMR spectra were recorded exclusively with composite pulse decoupling. Chemical shifts were referenced to $\delta_{TMS}$ = 0.00 ppm ($^{1}$H, $^{13}$C). Chemical shifts ($\delta$) are reported in ppm and coupling constants ($J$) are reported in Hz.

*Mass spectrometry*

Molecular masses were measured on a Advion expression CMS.

*X-ray measurements*

Suitable single crystals were coated with Paratone-N oil or Fomblin Y25 PFPE oil, mounted using a glass fiber, and frozen in the cold nitrogen stream. X-ray diffraction data were collected at 100 K on a Rigaku Oxford Diffraction SuperNova diffractometer using Cu $K_{\alpha}$ radiation ($\lambda$ = 1.54184 Å) generated by micro-focus sources. Data reduction and absorption correction was performed using CrysAlisPro[30], respectively. Using Olex2[31], the structures were solved with SHELXT[32] by direct methods and refined with SHELXL[32] by least-square minimization against $F^2$ using first isotropic and later anisotropic thermal parameters for all non-hydrogen atoms. Hydrogen atoms were added to the structure models on calculated positions using the riding model. Images of the structures were produced with the Olex2 software.

*Thermal analysis*

Simultaneous thermal analysis (STA) was conducted with a *STA 8000* apparatus (Perkin Elmer) under helium gas flow (20 mL/min) and a heating rate of 10 K/min. The "decomposition" temperature was obtained from 5 % weight loss and is disclosed as $T_d$.

*Materials simulations*

Density functional theory (DFT) and time-dependent DFT (TD-DFT) simulations were performed using the Gaussian 16 and the ORCA 6.0 software packages[33,34]. The long-range corrected hybrid functional CAM-B3LYP[35] was used in combination with the 6-311G**[36,37] basis set, as this level of theory has shown to accurately predict qualitative trends for the electronic structure and the photophysical properties of organic materials – partially even at sufficient quantitative accuracy[15,38,39].

Where available, molecular structures from the crystal structures were used as initial structural guess. Additional conformations were tested to identify the molecule's global energy minimum. The accuracy of the geometry optimization was evaluated based on an analysis of the vibrational spectra for the relaxed structures. All results are based on the conformations found with the lowest total energy.

*Thin film fabrication for photophysical measurements*

1 mL of toluene was added to 15.79 mg of the emitter and 300 mg polystyrene ($M_W$ = 35000 g/mol). The mixture was stirred and gently heated until complete dissolution of polymer host and emitter. A volume of 150 µL of solution was used to produce uniform films via spin coating (novocontrol SCE-150) at a speed of 16 rps for 60 s. The film was annealed on a hot plate at 123 °C for 2:15 min. This procedure resulted in film thicknesses 3750 nm and 3900 nm with no significant trend among the different emitter materials.

*PLTs (programmable luminescent tags) fabrication*

1 mL of toluene was added to 15.79 mg of the emitter and 300 mg polystyrene ($M_W$ = 35000 g/mol). The mixture was stirred and gently heated until complete dissolution. For the oxygen blocking layer, 50 mg Exceval was dissolved in 1 mL of water:IPA 9:1 mixture at 120 °C.

For spin coating, a speed of 16 rps for 60 s and volumes of 200 µL of polymer:host and Exceval solutions were used to produce uniform films on a 1“x1“ quartz substrate. Before applying the oxygen barrier layer, the polymer:host films were annealed on a hot plate at 123 °C for 2:15 min. This procedure resulted in film thicknesses of the active layer of approx. 300 nm and the oxygen blocking layer of approx. 3700 nm with no significant trend among the different emitter materials.

*UV-vis absorption spectroscopy*

The absorbance of individual and blended films spin-coated onto quartz substrates was investigated using a Shimadzu SolidSpec-3700 UV-vis-NIR absorption spectrometer. The absorbance $A$ is calculated as $A = 1 - T - R$, with the transmission $T$ and the reflection $R$ obtained in an integrating sphere.

*Emission measurements*

Direct and delayed emission measurements were performed using a CAS 140CTS from Instrument Systems and triggered 275 nm (Thorlabs, M275L4) and 300 nm (Thorlabs, M300L4) LEDs, respectively. For automated data acquisition, the control software SweepMe! was used[40]. All measurements were performed in darkness under nitrogen or aerated conditions. Emission spectra of $10^{-5}$ M solutions of the compounds were measured with instruments mentioned above.

*Phosphorescence lifetime measurements*

The phosphorescence lifetime was determined using a silicon photodetector (PDA100A, Thorlabs). The decays were recorded and fitted using a suitable multiexponential fit. The procedure and details can be found in Refs. [41] and [42].

*Photoluminescence quantum yield*

The PLQY values were determined using the method proposed by de Mello et al.[43], improved by F. Fries and S. Reineke[44]. As excitation source, a 300 W xenon lamp combined with a monochromator (LOT Quantum Design MSH300) was used. The samples were placed in a calibrated integration sphere (Labsphere RTC-060-SF) and the spectra acquired with an array spectrometer (CAS 140CT, Instrument Systems).

*Activation Curves*

The activation curves of PLTs were determined using an experimental setup presented recently.[15] The activation behavior of PLTs was recorded under UV-illumination using a CMOS camera (acA1920-40uc, Basler) with focusing lenses (HF25XA-5M, Fujifilm). A 450 nm long-pass filter (FELH0450, Thorlabs) was placed in front of the lens to prevent UV overexposure of the recorded images. Two identical 280 nm (M280L6, Thorlabs) LEDs were used as the excitation source, illuminating the PLTs approximately homogeneously in continuous-wave mode. The mean intensity of the PLT was calculated for each recorded frame. The procedure was controlled using the SweepMe! measurement software[40]. For deactivation, the PLTs were placed on a hot plate at 90 °C for 30 minutes and then left to rest at room temperature for a further 30 min before the next measurement cycle.

*Activation Time and Dose*

The activation time $t_{\mathrm{act}}$ of a PLT was directly extracted from the first activation curve. The offset of this curve can be assigned to the fluorescence intensity $I_{\mathrm{F}}$, while the maximum detected intensity is attributed to the overall photoluminescence $I_{\mathrm{L}}$. The phosphorescent intensity is calculated as $I_{\mathrm{P}} = I_{\mathrm{L}} - I_{\mathrm{F}}$, and the intensity for reading out the activation time is given as $I(t_{\mathrm{act}}) = \frac{I_{\mathrm{P}}}{2} + I_{\mathrm{F}}$. The activation dose ($D_{\mathrm{act}}$) was calculated as $D_{\mathrm{act}} = I_0 \cdot t_{\mathrm{act}}$, with incident UV irradiance ($I_0$) measured using a calibrated spectrometer (CAS 140CTS from Instrument Systems) at the sample position.

## S2. Synthesis

### S2.1 Suzuki-Miyaura cross-coupling reactions for D-A compounds

The target compounds were synthesized *via* Suzuki-Miyaura cross-coupling reactions by using bromide precursor (1 eq.) phenoxathiin-4-boronic acid (1.2 eq.), aqueous potassium carbonate (3.0 eq.) as a base and tetrakis(triphenylphosphine)palladium(0) (0.03 eq.) as catalyst. All dry reagents were added to a Schlenk flask and the atmosphere was purged with nitrogen for 15 min. After that dry tetrahydrofuran (20 ml) was added and, additionally, a solution of potassium carbonate (3 eq.) dissolved in 3 ml of water was added dropwise. The reaction mixture was heated up to 70 °C and stirred for 24 hours under nitrogen atmosphere. After cooling down, the mixture was poured into water. Dichloromethane was used for extraction. The organic phase was washed with brine and afterwards dried with sodium sulfate ($Na_2SO_4$). The solvent was evaporated. The crude product was purified by column chromatography (hexane/ethyl acetate (7/1) as eluent) and recrystalized from the eluent.

### S2.1.1 Synthesis and verification of Py-PX

3-(phenoxathiin-4-yl)pyridine (**Py-PX**) was synthesized in accordance with the general procedure from 3-bromopyridine (0.50 g, 3.16 mmol), phenoxathiin-4-boronic acid (0.93 g, 3.81 mmol), tetrakis(triphenylphosphine)palladium(0) (0.11 g, 0.095 mmol) and potassium carbonate (1.31 g, 9.49 mmol). The pure product was isolated as white crystals with 93% yield (0.82 g, .2.96 mmol).

$^1$H NMR (500 MHz, $CD_2Cl_2$) δ 8.81 (s, 1H), 8.65 (d, *J* = 4.6 Hz, 1H), 7.94 – 7.91 (m, 1H), 7.46 (dd, *J* = 7.8, 4.9 Hz, 1H), 7.27 – 7.16 (m, 5H), 7.10 (t, *J* = 7.5 Hz, 1H), 6.95 (d, *J* = 8.0 Hz, 1H).
$^{13}$C NMR (126 MHz, $CD_2Cl_2$) δ 152.36, 150.12, 149.36, 148.61, 136.64, 132.80, 129.12, 128.19, 127.91, 126.92, 126.82, 124.96, 124.69, 122.99, 121.95, 120.86, 117.60.
MS (m/z): calculated for $C_{17}H_{11}NOS$ $[M]^+$ = 277.3, found $[M]^+$ = 278.0

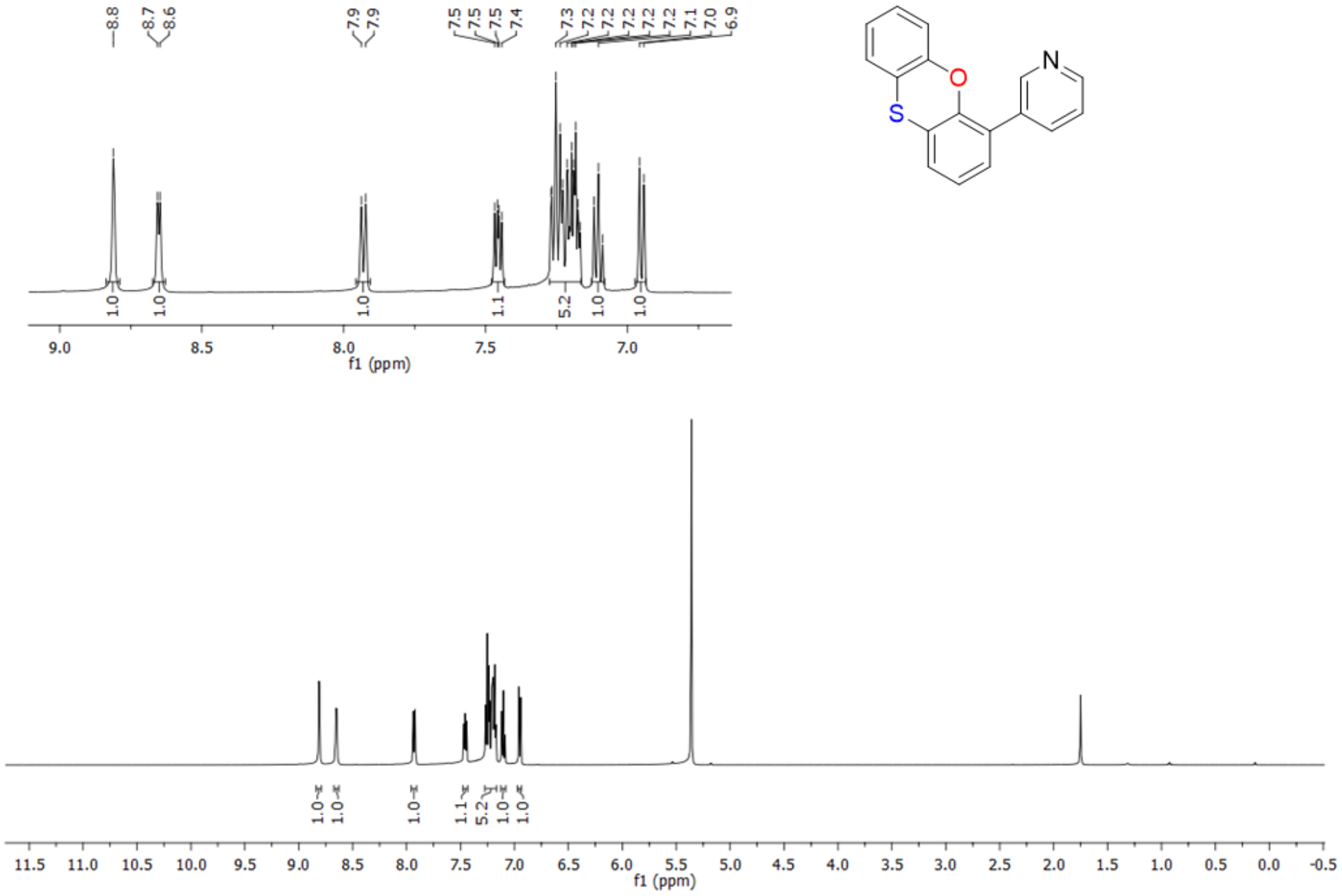

**Figure S1.** $^1$H NMR spectrum of **Py-PX** in $CD_2Cl_2$.

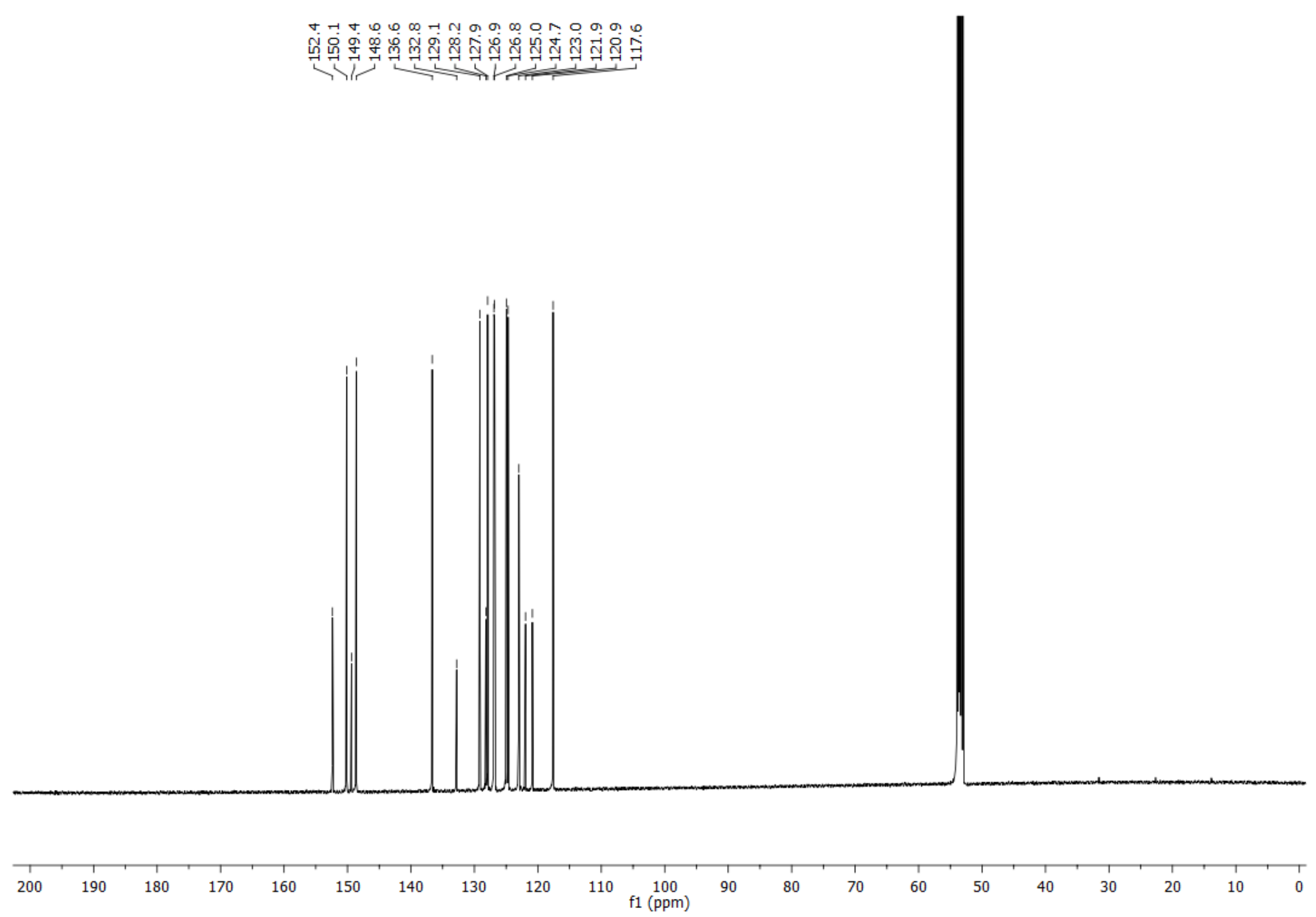

**Figure S2.** $^{13}C$ NMR spectrum of **Py-PX** in $CD_2Cl_2$.

### S2.1.2 Synthesis and verification of BP-PX

(4-(phenoxathiin-4-yl)phenyl)(phenyl)methanone (**BP-PX**) was synthesized in accordance with the general procedure from 4-brombenzophenon (0.50 g, 1.92 mmol), phenoxathiin-4-boronic acid (0.56 g, 2.29 mmol), tetrakis(triphenylphosphine)palladium(0) (0.066 g, 0.057 mmol) and potassium carbonate (1.74 g, 12.6 mmol). The pure product was isolated as white crystals with 50% yield (0.37 g, 0.97 mmol).

$^{1}$H NMR (500 MHz, $CD_2Cl_2$) δ 7.96 (d, *J* = 8.2 Hz, 2H), 7.91 (d, *J* = 7.2 Hz, 2H), 7.73 (d, *J* = 8.3 Hz, 2H), 7.68 (t, *J* = 7.4 Hz, 1H), 7.58 (t, *J* = 7.6 Hz, 2H), 7.31 (dd, *J* = 7.5, 1.6 Hz, 1H), 7.24 (td, *J* = 8.2, 1.5 Hz, 2H), 7.19 (dd, *J* = 8.9, 6.2 Hz, 2H), 7.11 (td, *J* = 7.5, 1.1 Hz, 1H), 6.99 (dd, *J* = 8.0, 0.9 Hz, 1H).
$^{13}$C NMR (126 MHz, $CD_2Cl_2$) δ 196.00, 152.49, 149.28, 141.33, 137.74, 136.51, 132.36, 130.55, 129.92, 129.89, 129.46, 129.23, 128.31, 127.90, 126.84, 124.94, 124.61, 122.03, 120.99, 117.58.
MS (m/z): calculated for $C_{25}H_{16}O_2S$ $[M]^+$ = 380.46, found $[M]^+$ = 381.0.

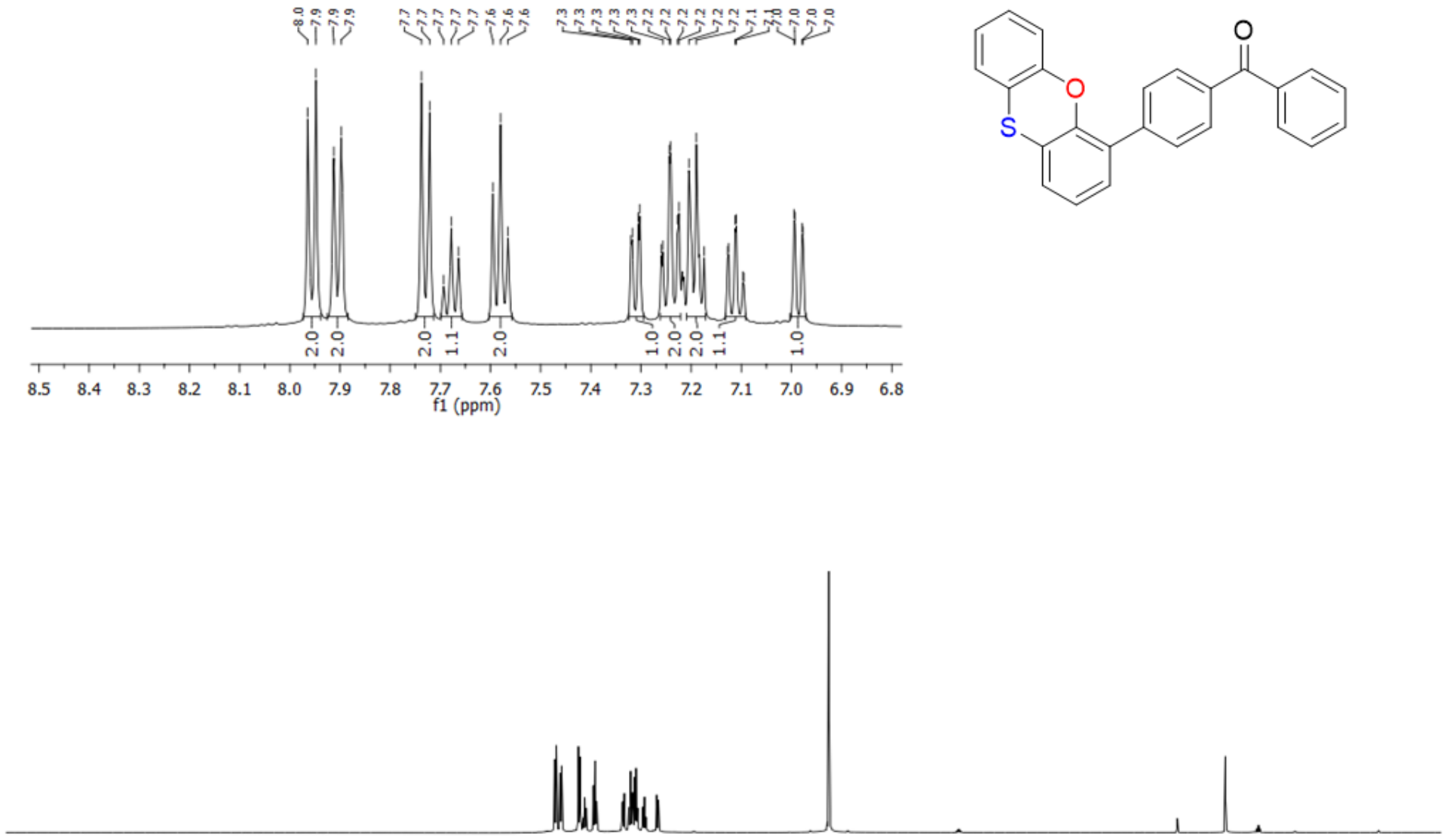

**Figure S3.** $^{1}$H NMR spectrum of **BP-PX** in $CD_2Cl_2$.

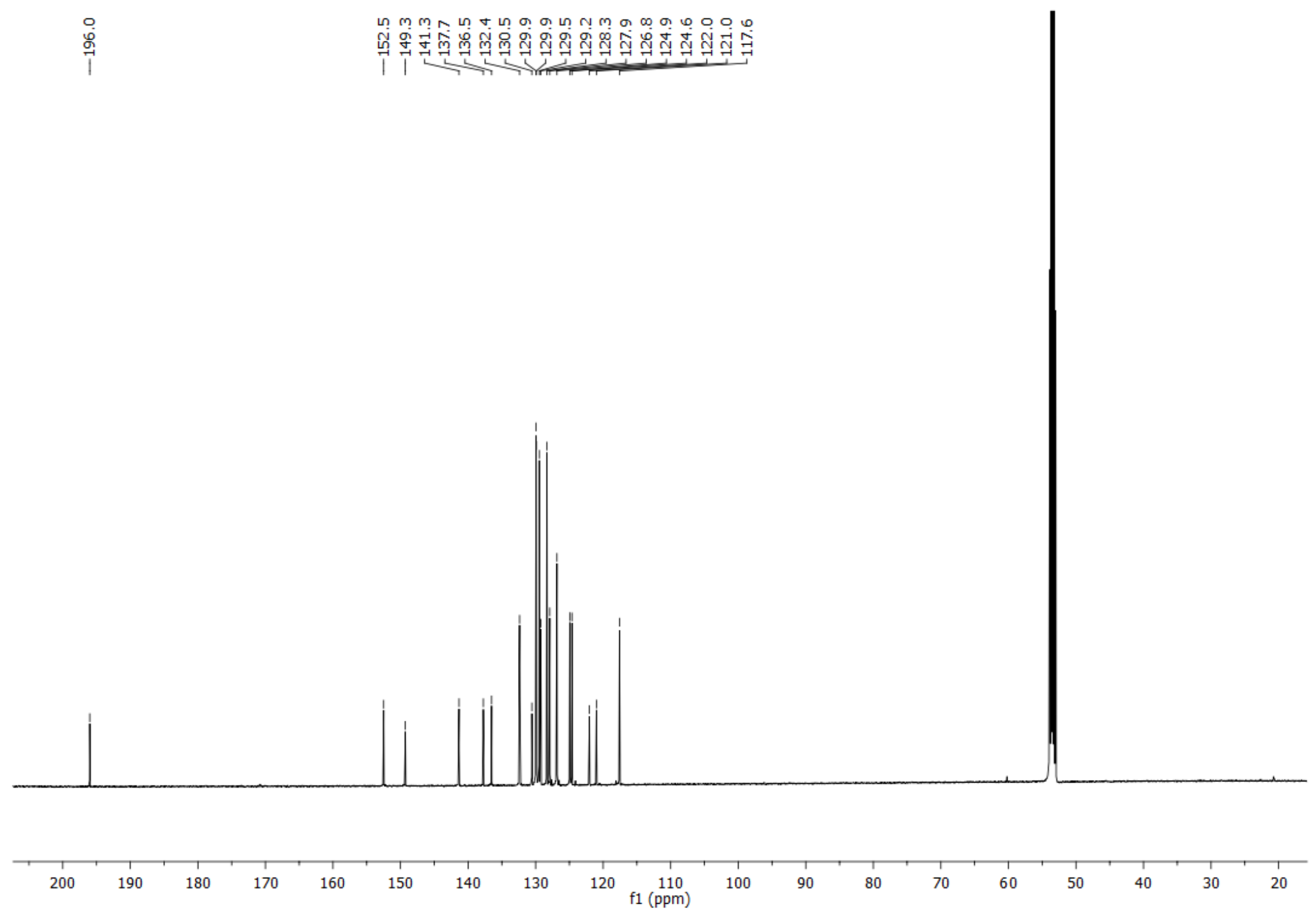

**Figure S4.** $^{13}$C NMR spectrum of **BP-PX** in $CD_2Cl_2$.

### S2.2 Suzuki-Miyaura cross-coupling reactions for D-A-D compounds

The target compounds were synthesized *via* Suzuki-Miyaura cross-coupling reactions by using bromide precursor (1 eq.) phenoxathiin-4-boronic acid (2.2 eq.), aqueous potassium carbonate (6.0 eq.) as a base, and tetrakis(triphenylphosphine)palladium(0) (0.05 eq.) as catalyst. All dry reagents were added to a Schlenk flask and the atmosphere was purged with nitrogen for 15 min. After that, dry tetrahydrofuran (20 ml) was added and, additionally, a solution of potassium carbonate (6 eq.) dissolved in 5 ml of water was added dropwise. The reaction mixture was heated up to 70 °C and stirred for 24 hours under nitrogen atmosphere. After cooling down, the mixture was poured into water. Dichloromethane was used for extraction. The organic phase was washed with brine and afterwards dried with sodium sulfate ($Na_2SO_4$). The solvent was evaporated. The crude product was purified by column chromatography (hexane/ethyl acetate (7/1) as eluent) and recrystalized from the eluent.

### S2.2.1 Synthesis and verification of Py-2PX

3,5-di(phenoxathiin-4-yl)pyridine (**Py-2PX**) was synthesized was synthesized in accordance with the general procedure from 3,5-dibromopyridine (0.50 g, 2.11 mmol), phenoxathiin-4-boronic acid (1.13 g, 4.63 mmol), tetrakis(triphenylphosphine)palladium(0) (0.12 g, 0.10 mmol) and potassium carbonate (1.75 g, 12.68 mmol). The pure product was isolated as white crystals with 49% yield (0.49 g, 1.02 mmol).

$^{1}$H NMR (500 MHz, $CD_2Cl_2$) δ 8.85 (d, *J* = 2.0 Hz, 2H), 8.13 (s, 1H), 7.35 (dd, *J* = 7.5, 1.6 Hz, 2H), 7.27 (dd, *J* = 7.7, 1.6 Hz, 2H), 7.22 (t, *J* = 7.6 Hz, 4H), 7.16 (td, *J* = 7.7, 1.5 Hz, 2H), 7.10 (td, *J* = 7.5, 1.1 Hz, 2H), 6.98 (dd, *J* = 7.9, 0.9 Hz, 2H).

$^{13}$C NMR (126 MHz, $CD_2Cl_2$) δ 152.31, 149.48, 149.02, 137.24, 132.36, 129.24, 128.03, 127.93, 127.05, 126.82, 124.98, 124.77, 122.00, 120.84, 117.71.

MS (m/z): calculated for $C_{29}H_{17}NO_2S_2$ $[M]^+$ = 475.58, found $[M]^+$ = 476.0.

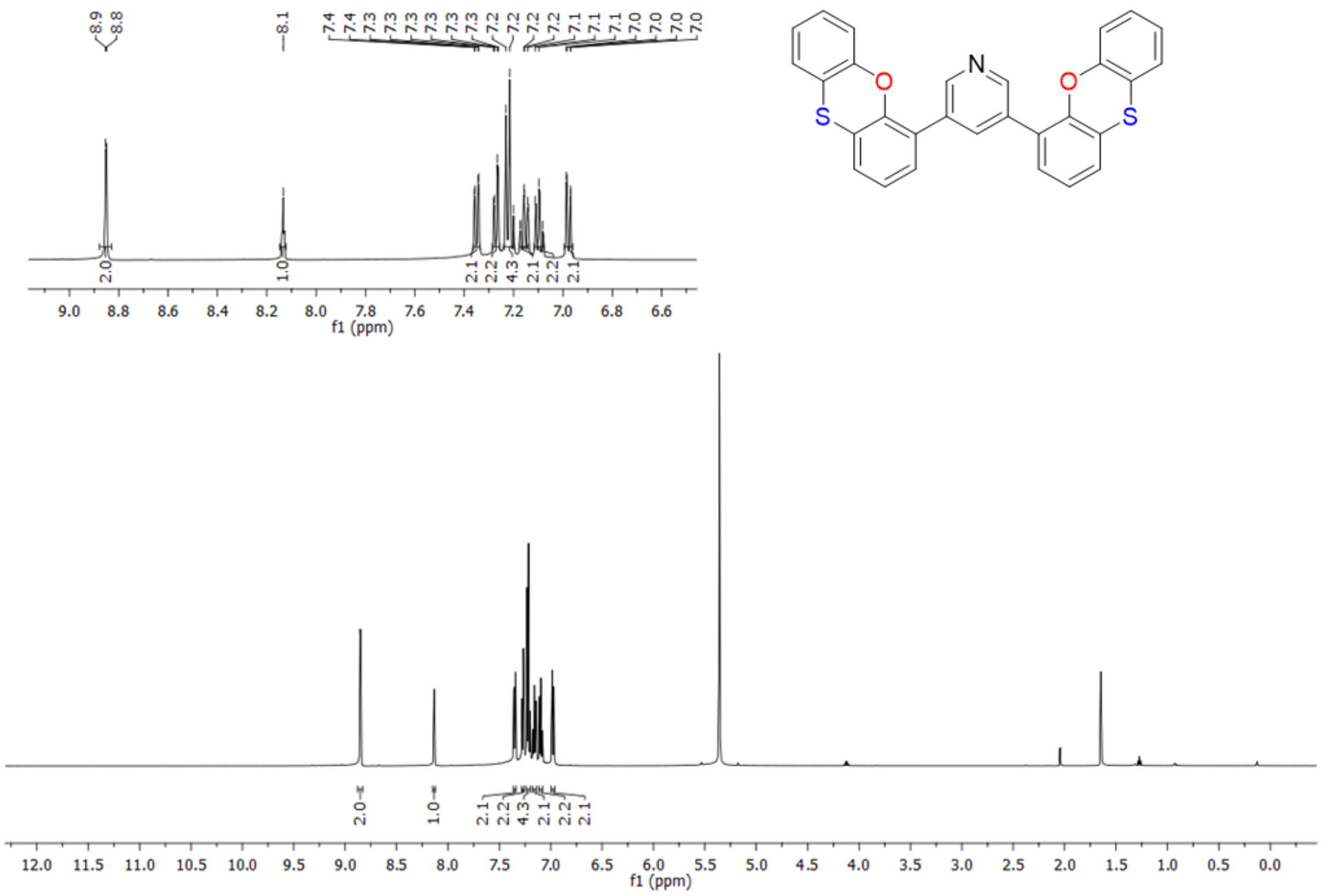

**Figure S5.** $^{1}$H NMR spectrum of **Py-2PX** in $CD_2Cl_2$.

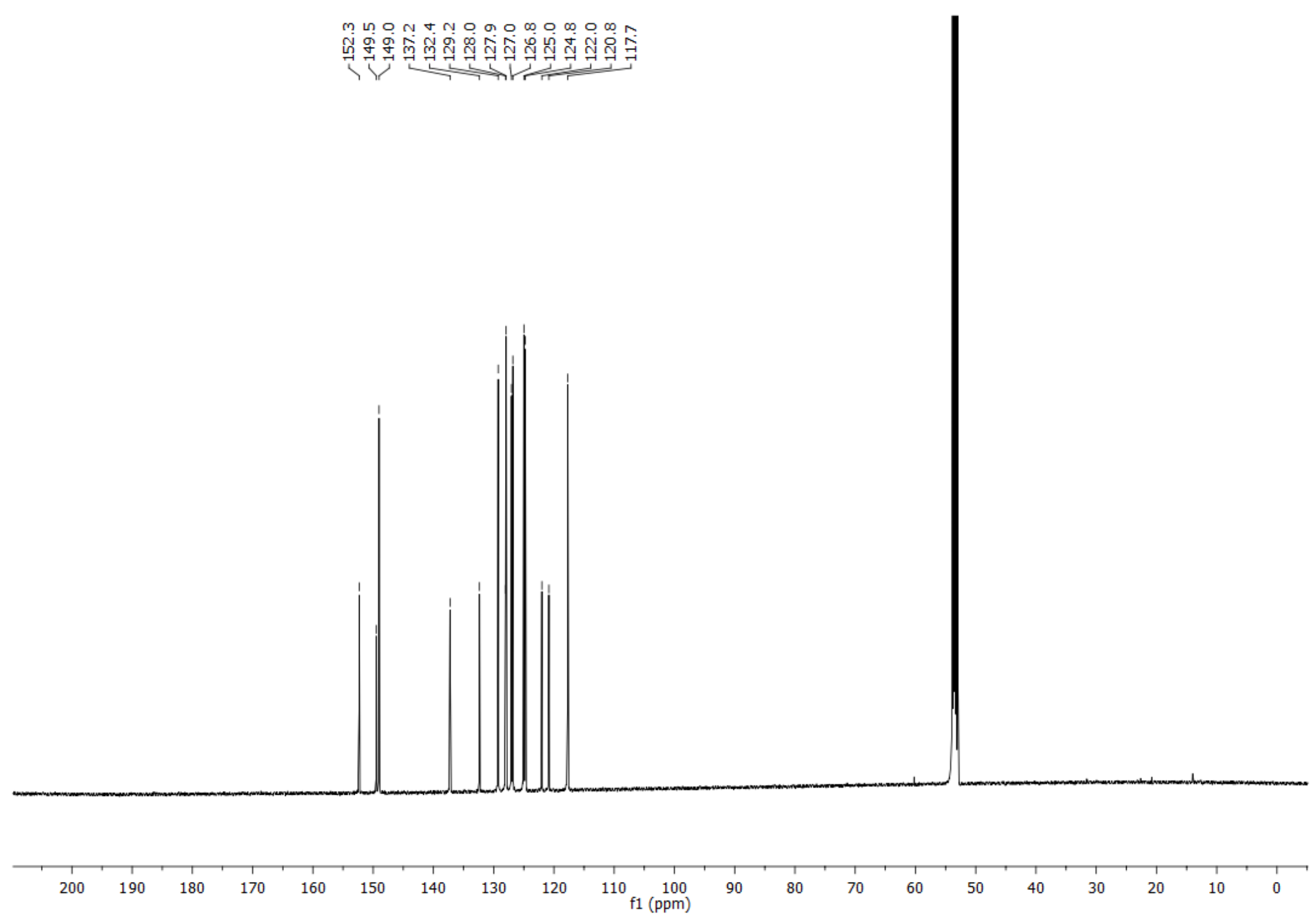

**Figure S6.** $^{13}$C NMR spectrum of **Py-2PX** in $CD_2Cl_2$.

### S2.2.2 Synthesis and verification of BP-2PX

Synthesis of bis(4-(phenoxathiin-4-yl)phenyl)methanone (**BP-2PX**) was synthesized according to the general procedure from 4,4′-dibrombenzophenon (0.50 g, 1.47 mmol), phenoxathiin-4-boronic acid (0.79 g, 3.24 mmol), tetrakis(triphenylphosphine)palladium(0) (0.085 g, 0.07 mmol) and potassium carbonate (1.22 g, 8.84 mmol). The pure product was isolated as white crystals with 40% yield (0.34 g, 0.59 mmol).

1H NMR (500 MHz, $CD_2Cl_2$) δ 8.03 (d, J = 8.3 Hz, 4H), 7.77 (d, J = 8.3 Hz, 4H), 7.33 (dd, J = 7.5, 1.6 Hz, 2H), 7.28 – 7.24 (m, 3H), 7.24 – 7.18 (m, 5H), 7.11 (dd, J = 10.7, 4.3 Hz, 2H), 7.02 – 6.99 (m, 2H).

$^{13}$C NMR (126 MHz, $CD_2Cl_2$) δ 195.66, 152.51, 149.30, 141.36, 136.62, 130.57, 129.89, 129.52, 129.25, 127.91, 126.86, 126.84, 124.95, 124.63, 122.04, 121.01, 117.59.

MS (m/z): calculated for $C_{37}H_{22}O_3S_2$ $[M]^+$ = 578.70, found $[M]^+$ = 579.0.

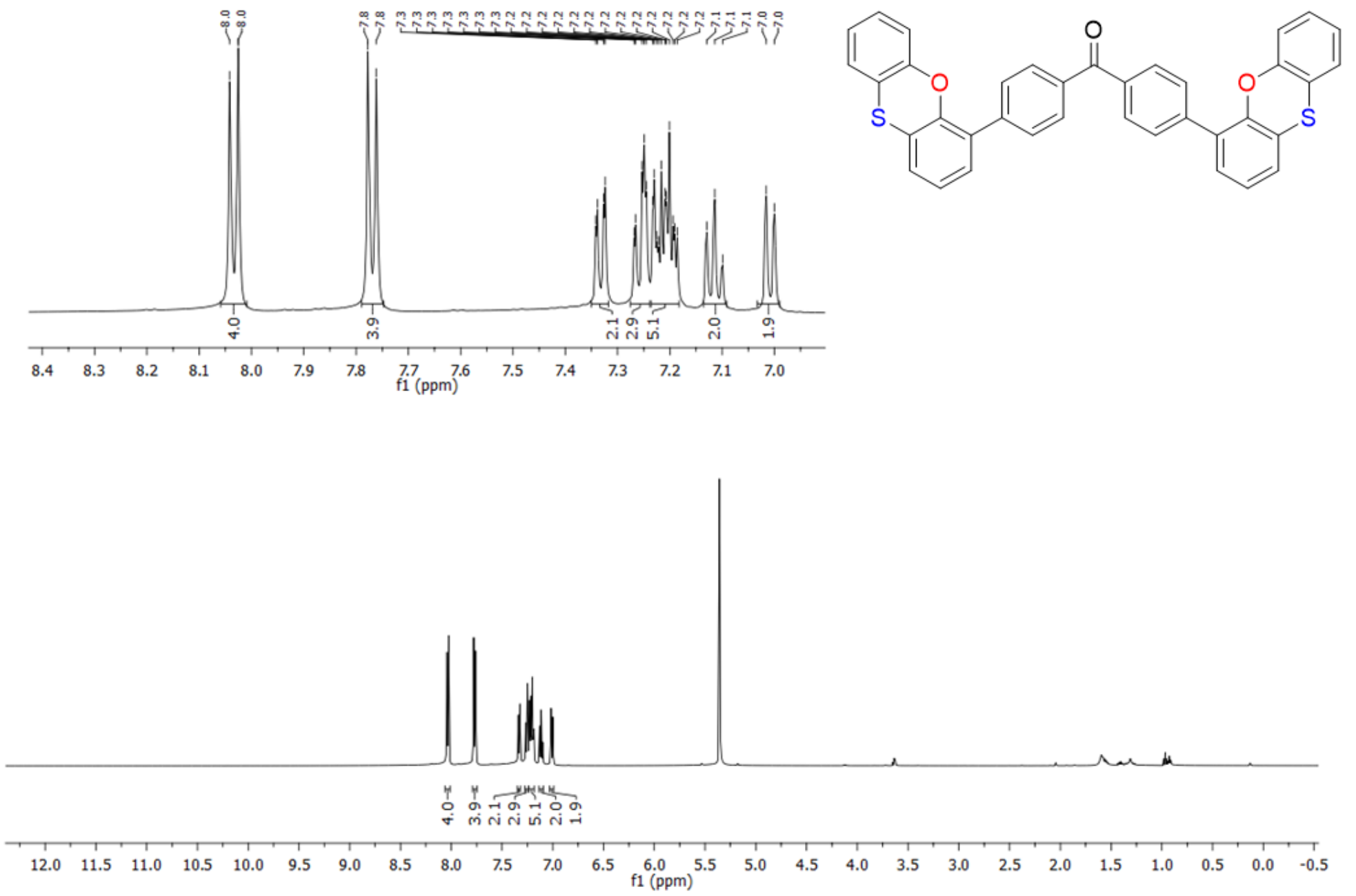

**Figure S7.** $^1$H NMR spectrum of **BP-2PX** in $CD_2Cl_2$.

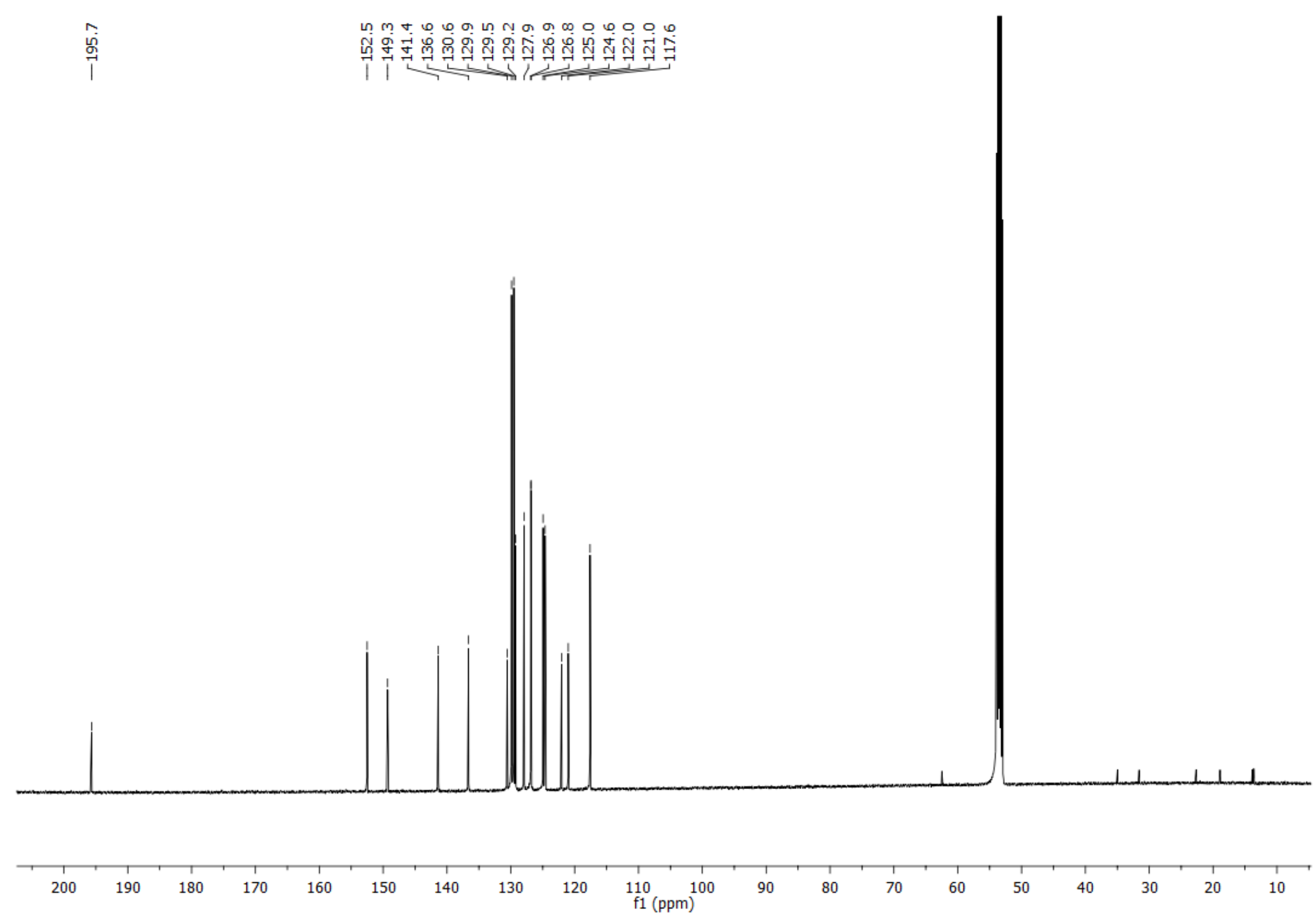

**Figure S8.** $^{13}C$ NMR spectrum of **BP-2PX** in $CD_2Cl_2$.

### S2.3 General procedure of synthesis for asymmetrical compounds

The asymmetrical compounds were obtained by two-step synthesis *via* Suzuki-Miyaura cross-coupling reactions.

#### S2.3.1 Synthesis and verification of Br-Py-TA

3-Bromo-5-(thianthren-1-yl)pyridine was synthesized by using 3,5-dibromopyridine (1 eq., 1.00 g, 4.22 mmol), thianthrene-1-boronic acid (1 eq., 1.1 g, 4.23 mmol), tetrakis(triphenylphosphine)palladium(0) (0.03 eq., 0.15 g, 0.13 mmol) and potassium carbonate (3 eq., 1.75 g, 12.68 mmol). All dry reagents were added to a Schlenk flask and the atmosphere was purged with nitrogen for 15 min. After that dry tetrahydrofuran (20 ml) was added and, additionally, a solution of potassium carbonate dissolved in 5 ml of water was added dropwise. The reaction mixture was heated up to 70 °C and stirred for 24 hours under nitrogen atmosphere. After cooling down, the mixture was poured into water. Dichloromethane was used for extraction. The organic phase was washed with brine and afterwards dried with sodium sulfate ($Na_2SO_4$). The solvent was evaporated. The crude product was purified by column chromatography (hexane/ethyl acetate (5/1) as eluent) and recrystalized from the eluent. The pure product was isolated as white crystals with 35 % yield (0.55 g, 1.48 mmol).

$^1$H NMR (500 MHz, $CD_2Cl_2$) δ 8.77 (s, 1H), 8.63 (s, 1H), 7.96 (s, 1H), 7.64 (d, *J* = 7.8 Hz, 1H), 7.56 (d, *J* = 7.6 Hz, 1H), 7.45 (d, *J* = 7.6 Hz, 1H), 7.39 (t, *J* = 7.7 Hz, 1H), 7.34 – 7.27 (m, 3H).
$^{13}$C NMR (126 MHz, $CD_2Cl_2$) δ 149.88, 148.26, 139.14, 137.27, 137.23, 136.65, 136.04, 135.10, 135.03, 129.25, 128.91, 128.58, 128.12, 127.84, 127.56, 120.10.
MS (m/z): calculated for $C_{17}H_{10}BrNS_2$ $[M]^+$ = 372.30, found $[M]^+$ = 373.9.

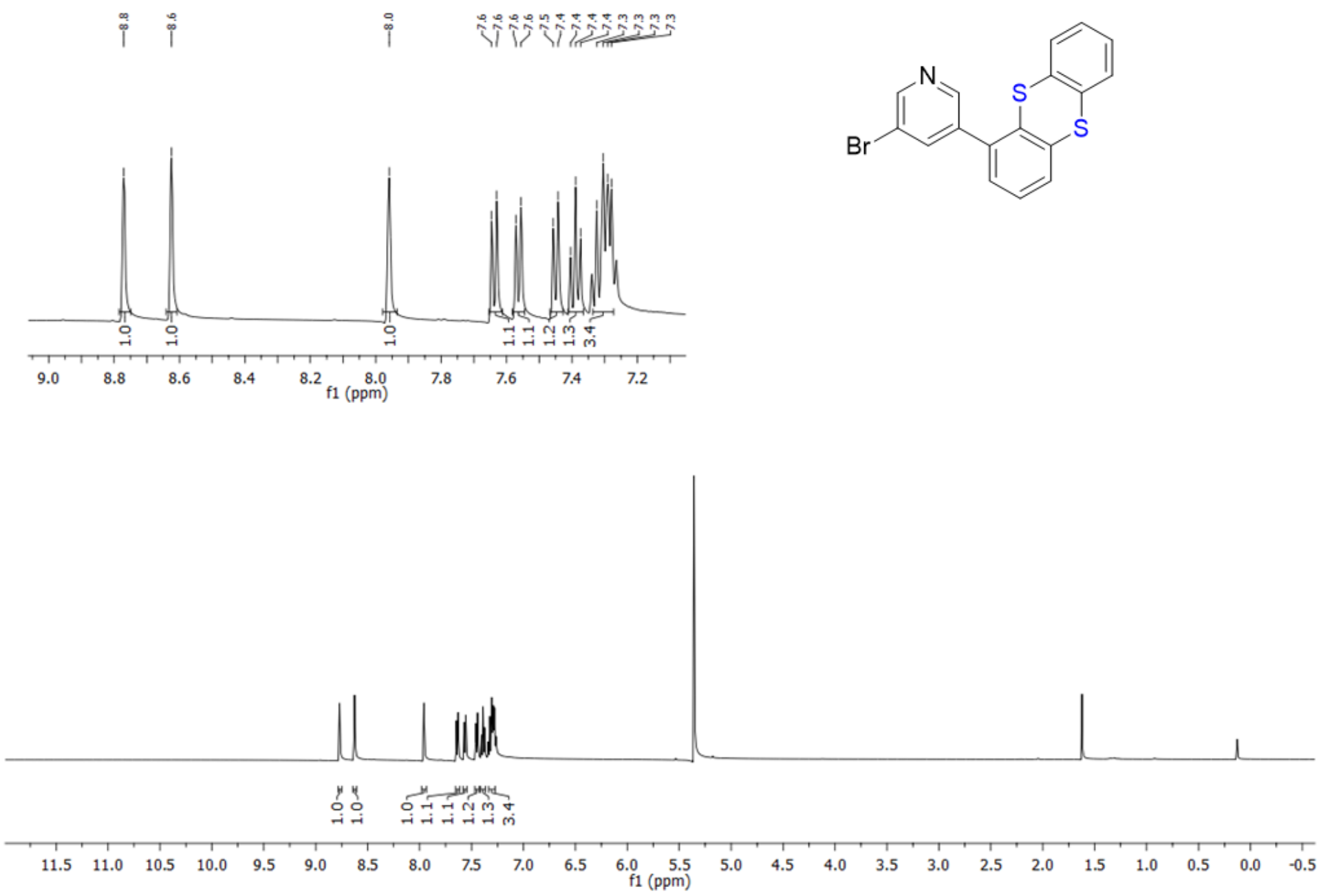

**Figure S9.** $^{1}H$ NMR spectrum of **Br-Py-TA** in $CD_2Cl_2$.

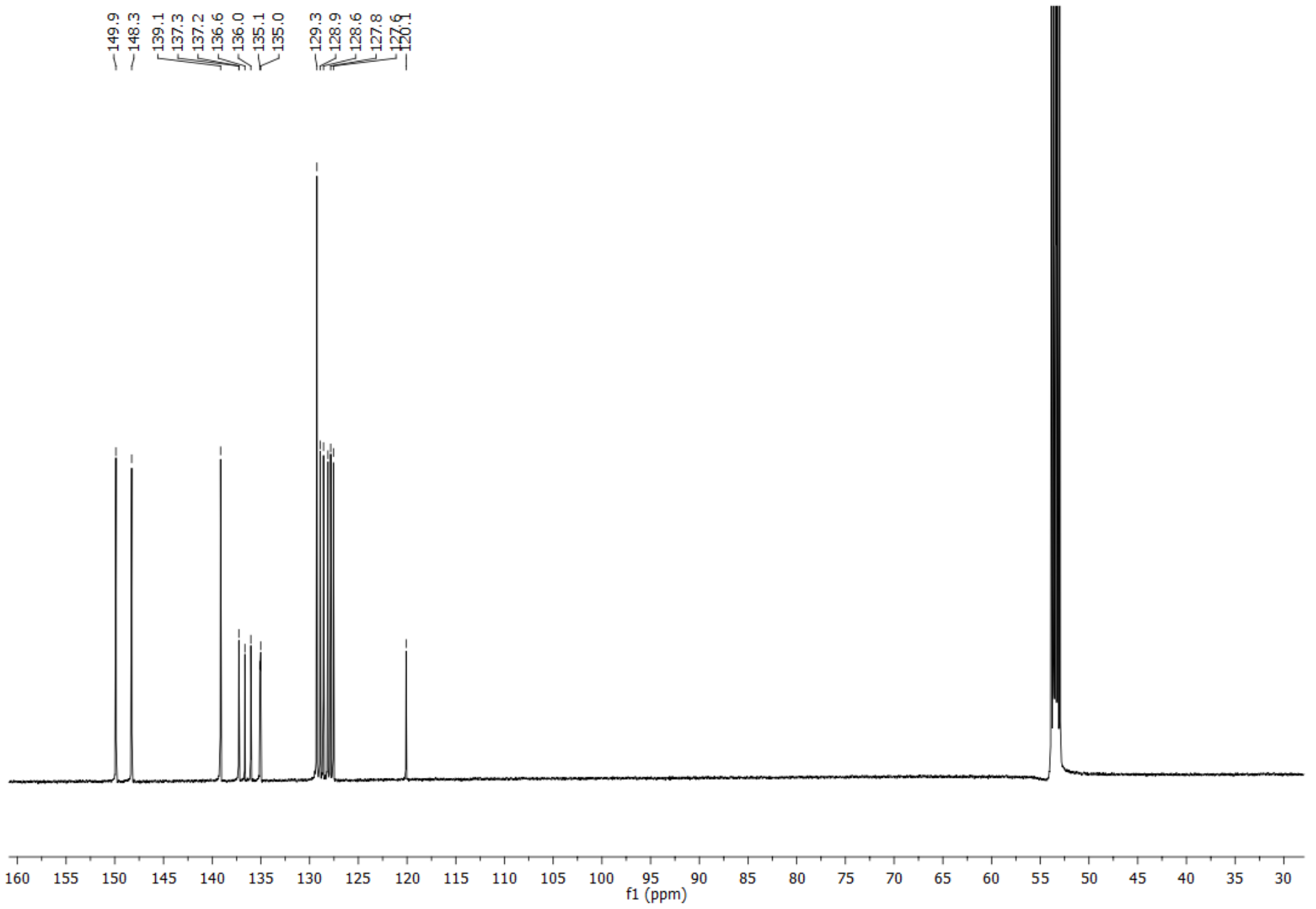

**Figure S10.** $^{13}C$ NMR spectrum of **Br-Py-TA** in $CD_2Cl_2$.

### S2.3.2 Synthesis and verification of PX-Py-TA

3-(Phenoxathiin-4-yl)-5-(thianthren-1-yl)pyridine (**PX-Py-TA**) was synthesized by using 3-bromo-5-(thianthren-1-yl)pyridine (Br-Py-TA) (1 eq., 0.4 g, 1.07 mmol), phenoxathiin-4-boronic acid (2.2 eq., 0.58 g, 2.38 mmol), tetrakis(triphenylphosphine)palladium(0) (0.05 eq., 0.06 g, 0.05 mmol) and potassium carbonate (6 eq., 0.89 g, 6.45 mmol). All dry reagents were added to a Schlenk flask and the atmosphere was purged with nitrogen for 15 min. After that dry tetrahydrofuran (20 ml) was added and, additionally, a solution of potassium carbonate dissolved in 5 ml of water was added dropwise. The reaction mixture was heated up to 70 °C and stirred for 24 hours under nitrogen atmosphere. After cooling down, the mixture was poured into water. Dichloromethane was used for extraction. The organic phase was washed with brine and afterwards dried with sodium sulfate ($Na_2SO_4$). The solvent was evaporated. The crude product was purified by column chromatography (hexane/ethyl acetate (5/1) as eluent) and recrystalized from the eluent. The pure product was isolated as white crystals with 65 % yield (0.34 g, 0.69 mmol).

$^1$H NMR (500 MHz, $CD_2Cl_2$) δ 8.89 (s, 1H), 8.71 (s, 1H), 8.01 (s, 1H), 7.66 – 7.63 (m, 1H), 7.57 (d, $J$ = 7.7 Hz, 1H), 7.46 (d, $J$ = 7.6 Hz, 1H), 7.41 (d, $J$ = 6.2 Hz, 2H), 7.36 (d, $J$ = 7.5 Hz, 1H), 7.32 (t, $J$ = 7.5 Hz, 1H), 7.26 (t, $J$ = 6.8 Hz, 2H), 7.22 (t, $J$ = 7.3 Hz, 2H), 7.16 (t, $J$ = 7.6 Hz, 1H), 7.10 (t, $J$ = 7.5 Hz, 1H), 7.01 (d, $J$ = 8.0 Hz, 1H).

$^{13}$C NMR (126 MHz, $CD_2Cl_2$) δ 152.27, 149.45, 149.31, 148.68, 138.73, 137.41, 136.26, 136.14, 135.32, 135.28, 135.22, 132.20, 129.41, 129.24, 128.91, 128.87, 128.55, 128.03, 127.92, 127.81, 127.77, 127.51, 127.11, 126.82, 124.98, 124.77, 121.96, 120.79, 117.72.

MS (m/z): calculated for $C_{29}H_{17}NOS_3$ $[M]^+$ = 491.64, found $[M]^+$ = 491.9.

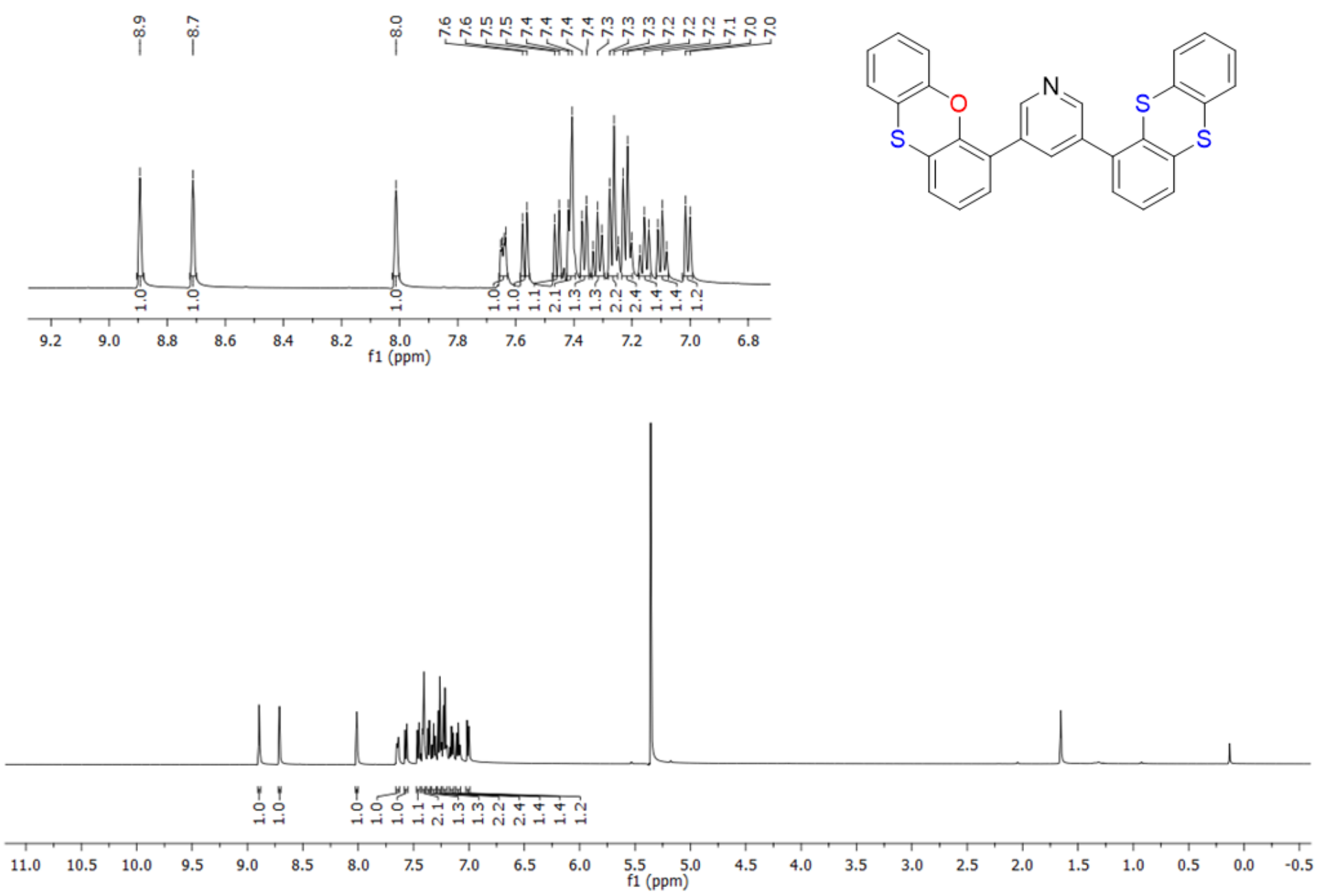

**Figure S11.** $^{1}H$ NMR spectrum of **PX-Py-TA** in $CD_2Cl_2$.

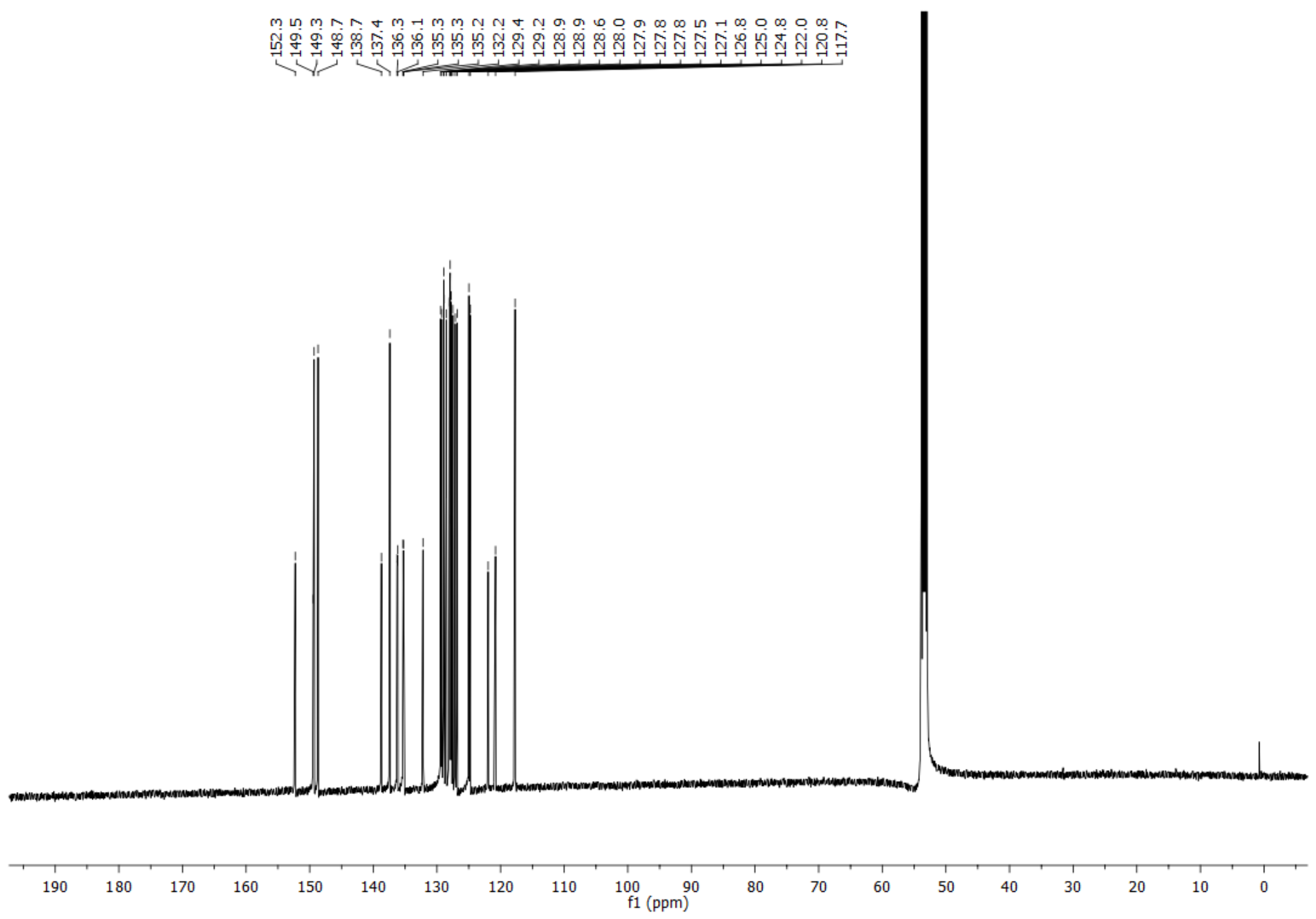

**Figure S12.** $^{13}C$ NMR spectrum of **PX-Py-TA** in $CD_2Cl_2$.

### S2.3.3 Synthesis and verification of Br-BP-TA

(4-Bromophenyl)(4-(thianthren-1-yl)phenyl)methanone was synthesized by using 4,4′-dibrombenzophenon (1 eq., 1.00 g, 2.94 mmol), thianthrene-1-boronic acid (1 eq., 0.76 g, 2.92 mmol), tetrakis(triphenylphosphine)palladium(0) (0.03 eq., 0.10 g, 0.09 mmol) and potassium carbonate (3 eq., 1.22 g, 8.84 mmol). All dry reagents were added to a Schlenk flask and the atmosphere was purged with nitrogen for 15 min. After that dry tetrahydrofuran (20 ml) was added and, additionally, a solution of potassium carbonate dissolved in 5 ml of water was added dropwise. The reaction mixture was heated up to 70 °C and stirred for 24 hours under nitrogen atmosphere. After cooling down, the mixture was poured into water. Dichloromethane was used for extraction. The organic phase was washed with brine and afterwards dried with sodium sulfate ($Na_2SO_4$). The solvent was evaporated. The crude product was purified by column chromatography (hexane/ethyl acetate (5/1) as eluent) and recrystalized from the eluent. The pure product was isolated as white crystals with 44 % yield (0.62 g, 1.30 mmol).

$^{1}$H NMR (500 MHz, $CD_2Cl_2$) δ 7.93 (d, *J* = 8.1 Hz, 2H), 7.80 (d, *J* = 8.4 Hz, 2H), 7.73 (d, *J* = 8.4 Hz, 2H), 7.61 (t, *J* = 8.2 Hz, 3H), 7.56 (d, *J* = 7.6 Hz, 1H), 7.43 (d, *J* = 7.6 Hz, 1H), 7.38 (t, *J* = 7.5 Hz, 1H), 7.34 (dd, *J* = 5.5, 4.5 Hz, 1H), 7.32 – 7.29 (m, 1H), 7.26 (t, *J* = 7.4 Hz, 1H).
$^{13}$C NMR (126 MHz, $CD_2Cl_2$) δ 194.86, 144.45, 141.33, 136.41, 136.38, 136.19, 135.96, 135.47, 134.78, 131.65, 131.53, 129.75, 129.57, 129.13, 128.78, 128.73, 128.52, 127.98, 127.72, 127.36, 127.32.
MS (m/z): calculated for $C_{25}H_{15}BrOS_2$ $[M]^+$ = 475.42, found $[M]^+$ = 476.9.

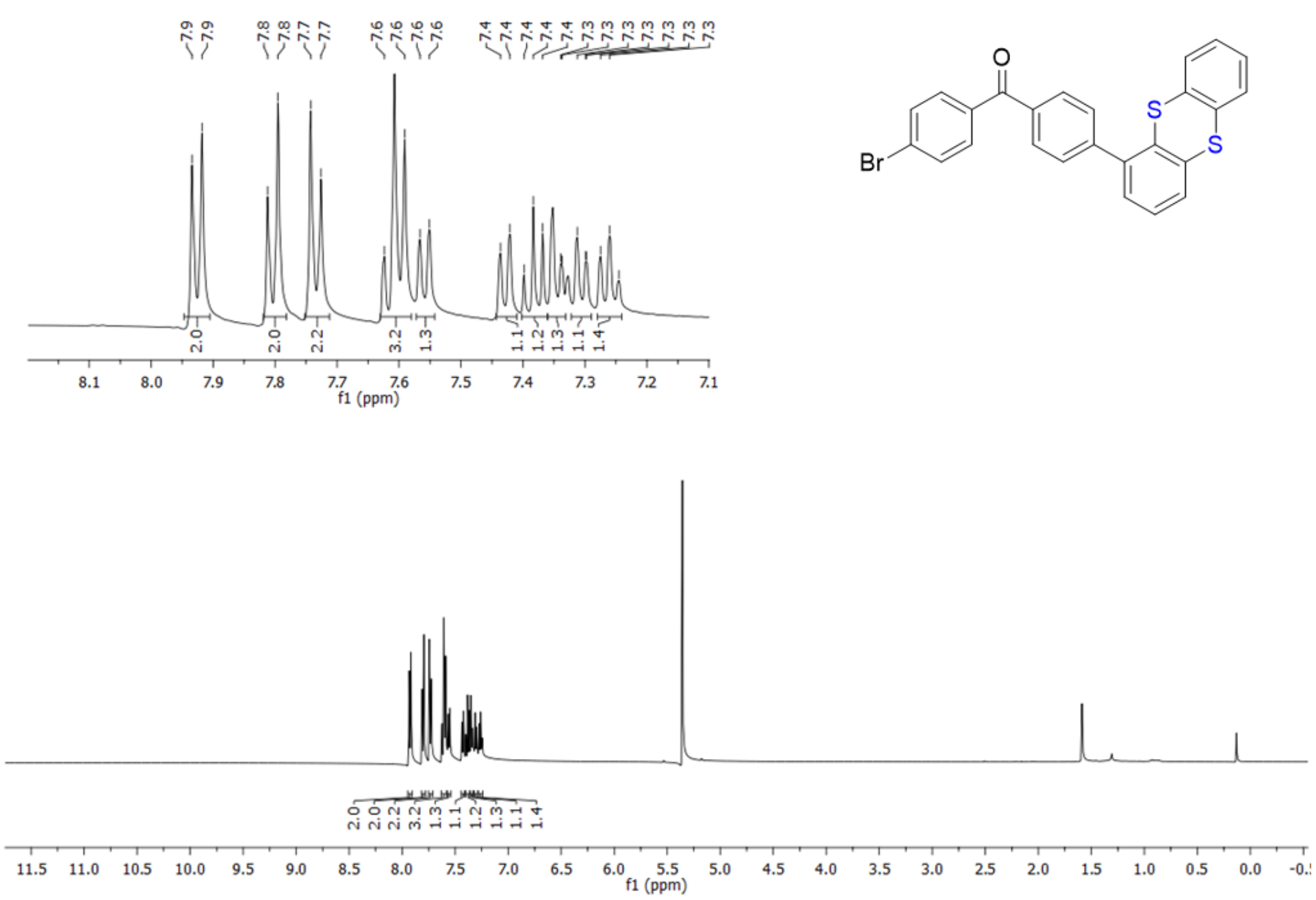

**Figure S13.** $^{1}H$ NMR spectrum of **Br-BP-TA** in $CD_2Cl_2$.

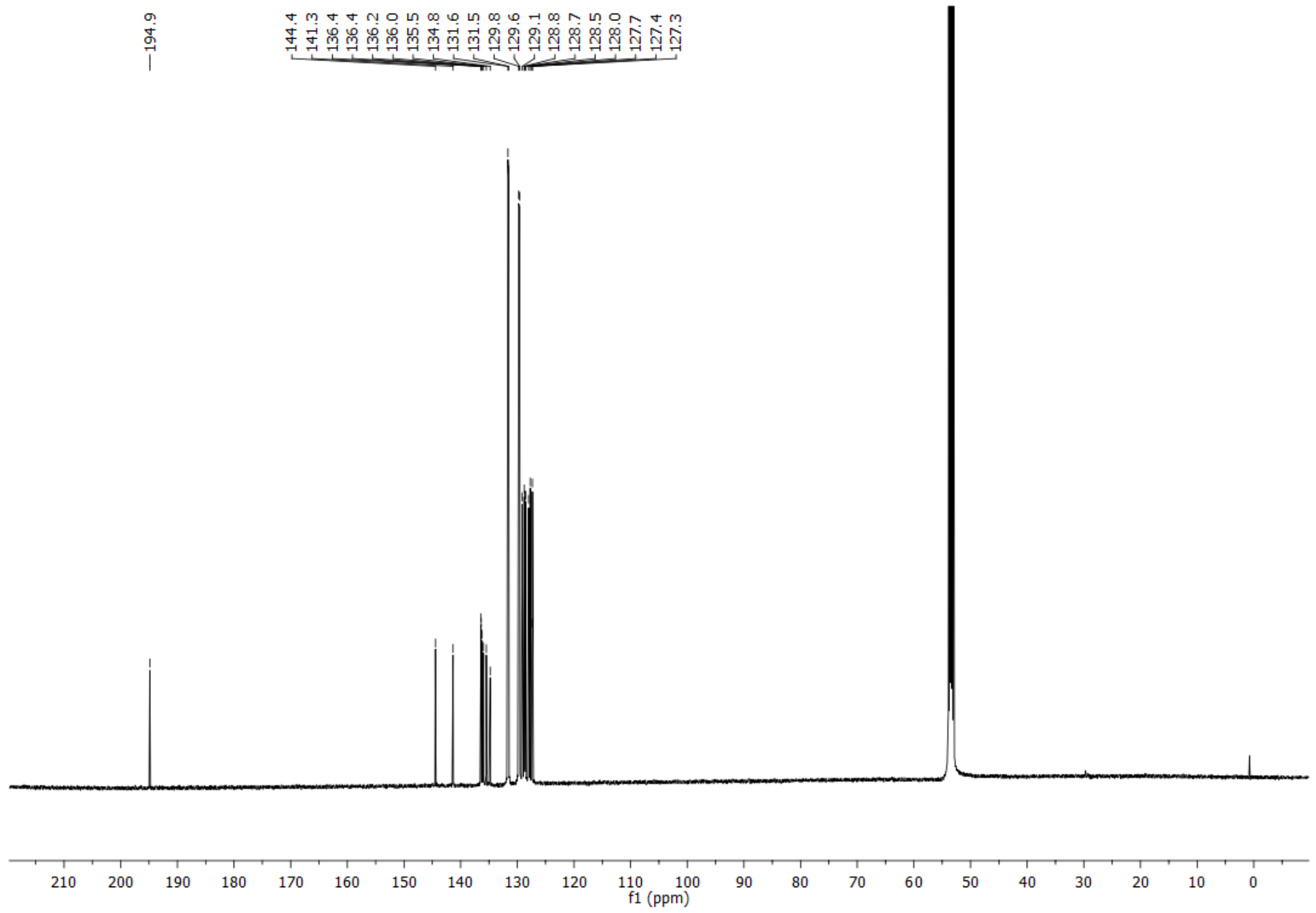

**Figure S14.** $^{13}C$ NMR spectrum of **Br-BP-TA** in $CD_2Cl_2$.

### S2.3.4 Synthesis and verification of PX-BP-TA

(4-(Phenoxathiin-4-yl)phenyl)(4-(thianthren-1-yl)phenyl)methanone (**PX-BP-TA**) was synthesized by using (4-bromophenyl)(4-(thianthren-1-yl)phenyl)methanone (Br-BP-TA) (1 eq., 0.29 g, 0.61 mmol), phenoxathiin-4-boronic acid (2.2 eq., 0.33 g, 1.35 mmol), tetrakis(triphenylphosphine)palladium(0) (0.05 eq., 0.04 g, 0.03 mmol) and potassium carbonate (6 eq., 0.51 g, 3.69 mmol). All dry reagents were added to a Schlenk flask and the atmosphere was purged with nitrogen for 15 min. After that dry tetrahydrofuran (20 ml) was added and, additionally, a solution of potassium carbonate dissolved in 5 ml of water was added dropwise. The reaction mixture was heated up to 70 °C and stirred for 24 hours under nitrogen atmosphere. After cooling down, the mixture was poured into water. Dichloromethane was used for extraction. The organic phase was washed with brine and afterwards dried with sodium sulfate ($Na_2SO_4$). The solvent was evaporated. The crude product was purified by column chromatography (hexane/ethyl acetate (5/1) as eluent) and recrystalized from the eluent. The pure product was isolated as white crystals with 46 % yield (0.17 g, 0.28 mmol).

$^{1}$H NMR (500 MHz, $CD_2Cl_2$) δ 8.04 (t, $J$ = 8.7 Hz, 4H), 7.78 (d, $J$ = 8.0 Hz, 2H), 7.63 (d, $J$ = 7.8 Hz, 3H), 7.57 (d, $J$ = 7.6 Hz, 1H), 7.45 (d, $J$ = 7.6 Hz, 1H), 7.42 – 7.36 (m, 2H), 7.32 (dd, $J$ = 11.8, 7.5 Hz, 2H), 7.29 – 7.24 (m, 2H), 7.24 – 7.18 (m, 3H), 7.12 (t, $J$ = 7.5 Hz, 1H), 7.01 (d, $J$ = 8.0 Hz, 1H).

$^{13}$C NMR (126 MHz, $CD_2Cl_2$) δ 195.56, 152.48, 149.28, 144.21, 141.47, 141.44, 136.95, 136.46, 136.21, 135.92, 135.53, 134.82, 130.54, 129.91, 129.85, 129.53, 129.51, 129.24, 129.17, 128.80, 128.69, 128.52, 127.97, 127.90, 127.71, 127.32, 126.86, 126.83, 124.94, 124.61, 122.03, 120.98, 117.59.

MS (m/z): calculated for $C_{37}H_{22}O_2S_3$ $[M]^+$ = 594.76, found $[M]^+$ = 595.0.

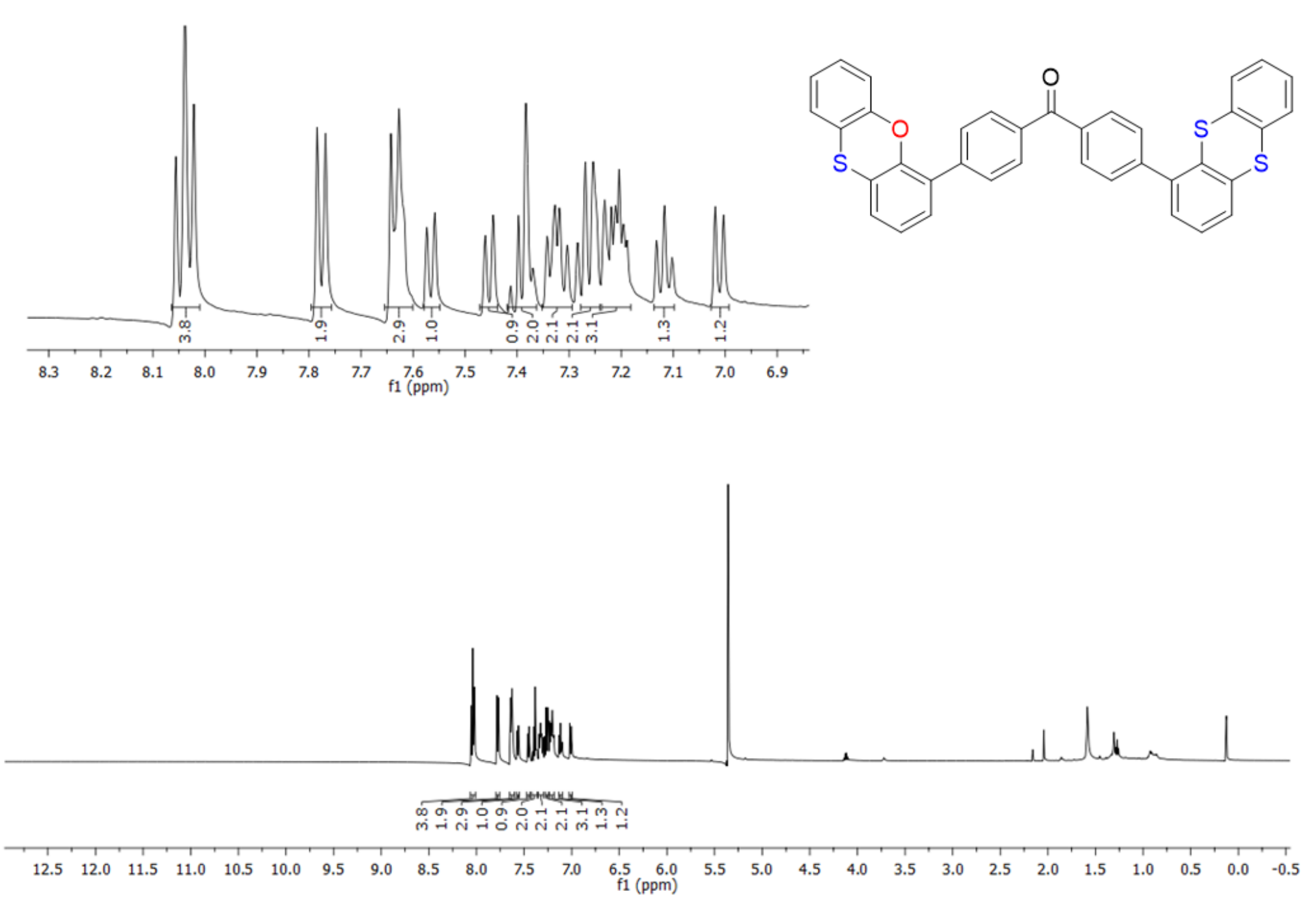

**Figure S15.** $^{1}$H NMR spectrum of **PX-BP-TA** in $CD_2Cl_2$.

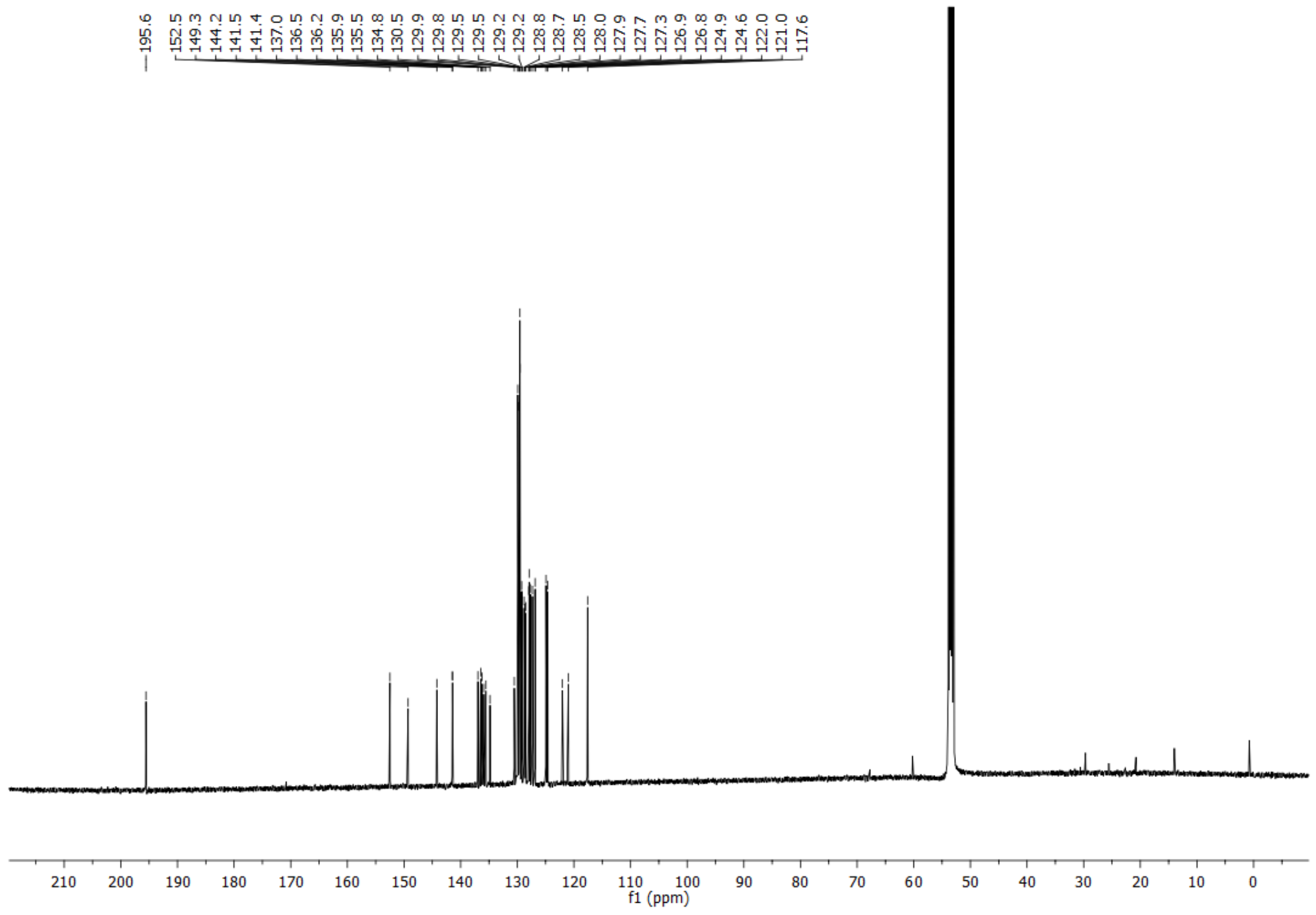

**Figure S16.** $^{13}$C NMR spectrum of **PX-BP-TA** in $CD_2Cl_2$.

## S3. Crystallographic details

| Compound | BP-PX | Py-2PX | Py-PX | TA-BP-PX |
|---|---|---|---|---|
| **Empirical formula** | $C_{25}H_{16}O_2S$ | $C_{29}H_{17}NO_2S_2$ | $C_{17}H_{11}NOS$ | $C_{37}H_{22}O_2S_3$ |
| **Formula weight** [g/mol] | 380.44 | 475.55 | 277.33 | 594.72 |
| **Temperature** [K] | 100.01(10) | 100.01(10) | 100.01(10) | 100.01(10) |
| **Crystal system** | orthorhombic | triclinic | monoclinic | triclinic |
| **Space group** | Pbca | P-1 | $P2_1/c$ | P-1 |
| **a** [Å] | 16.80434(11) | 10.7264(3) | 19.9304(2) | 10.25980(10) |
| **b** [Å] | 7.51749(5) | 10.7627(4) | 8.39720(10) | 11.0964(2) |
| **c** [Å] | 28.20831(17) | 11.0546(3) | 7.67840(10) | 14.09890(10) |
| **a** [°] | 90 | 98.870(3) | 90 | 105.3420(10) |
| **β** [°] | 90 | 91.017(3) | 96.3210(10) | 104.1480(10) |
| **γ** [°] | 90 | 119.820(4) | 90 | 108.0290(10) |
| **Volume** [$Å^3$] | 3563.45(4) | 1087.32(7) | 1277.24(3) | 1375.57(3) |
| **Z** | 8 | 2 | 4 | 2 |
| $\rho_{calc}$ [$g/cm^3$] | 1.418 | 1.453 | 1.442 | 1.436 |
| **μ** [$mm^{-1}$] | 1.759 | 2.454 | 2.187 | 2.742 |
| **F(000)** | 1584.0 | 492.0 | 576.0 | 616.0 |
| **Crystal size** [$mm^3$] | 0.26 × 0.13 × 0.11 | 0.31 × 0.16 × 0.04 | 0.22 × 0.19 × 0.12 | 0.238 × 0.169 × 0.049 |
| **Radiation** (λ in Å) | Cu Kα (λ = 1.54184) | Cu Kα (λ = 1.54184) | Cu Kα (λ = 1.54184) | Cu Kα (λ = 1.54184) |
| **2Θ range for data collection** [°] | 6.266 to 153.25 | 8.144 to 152.632 | 4.46 to 153.548 | 6.956 to 153.216 |
| **Index ranges** | $-21 \le h \le 20, -9 \le k \le 9, -28 \le l \le 35$ | $-13 \le h \le 13, -13 \le k \le 13, -7 \le l \le 13$ | $-25 \le h \le 24, -10 \le k \le 8, -9 \le l \le 7$ | $-10 \le h \le 12, -14 \le k \le 13, -17 \le l \le 17$ |
| **Reflections collected** | 43314 | 11114 | 12918 | 14893 |
| **Independent reflections** | 3755 [$R_{int}$ = 0.0434, $R_{sigma}$ = 0.0178] | 4509 [$R_{int}$ = 0.0203, $R_{sigma}$ = 0.0259] | 2668 [$R_{int}$ = 0.0183, $R_{sigma}$ = 0.0128] | 5704 [$R_{int}$ = 0.0260, $R_{sigma}$ = 0.0304] |
| **Data/restraints/parameters** | 3755/0/253 | 4509/0/307 | 2668/0/181 | 5704/19/479 |
| **Goodness-of-fit on $F^2$** | 1.056 | 1.062 | 1.078 | 1.045 |
| **Final R indexes** (I>=2σ (I)) | $R_1$ = 0.0337, $wR_2$ = 0.0902 | $R_1$ = 0.0287, $wR_2$ = 0.0758 | $R_1$ = 0.0289, $wR_2$ = 0.0759 | $R_1$ = 0.0334, $wR_2$ = 0.0843 |
| **Final R indexes** (all data) | $R_1$ = 0.0368, $wR_2$ = 0.0935 | $R_1$ = 0.0304, $wR_2$ = 0.0771 | $R_1$ = 0.0298, $wR_2$ = 0.0766 | $R_1$ = 0.0367, $wR_2$ = 0.0867 |
| **Largest diff. peak/hole** [e/$Å^3$] | 0.35/-0.33 | 0.30/-0.27 | 0.30/-0.25 | 0.32/-0.40 |
| **CCDC Number** | ### | ### | ### | ### |

## S4. Thermal properties

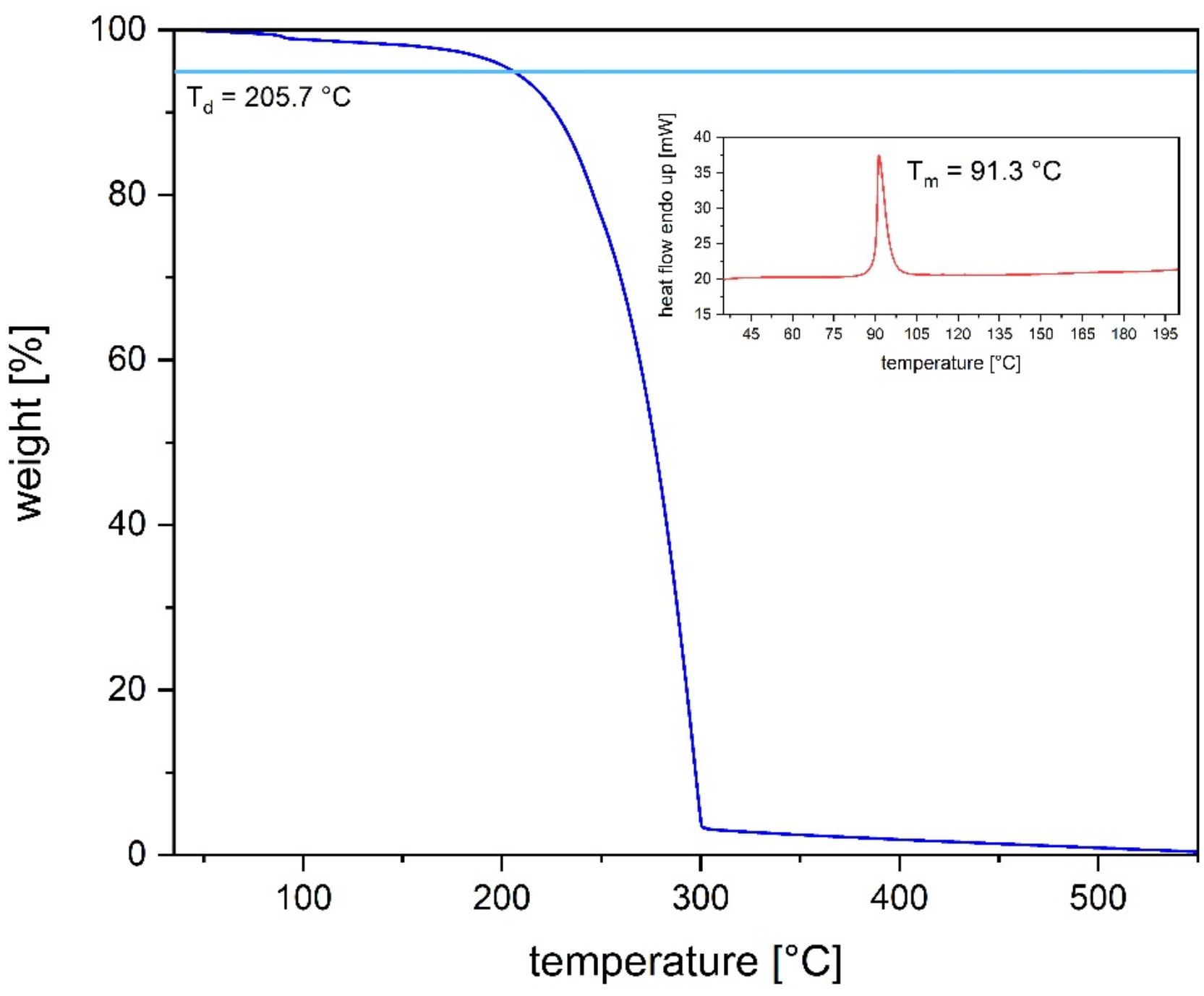

**Figure S17.** TGA and DSC (insert) results from simultaneous thermal analysis of **Py-PX** ($T_\mathrm{d}$ was obtained at 5% weight loss).

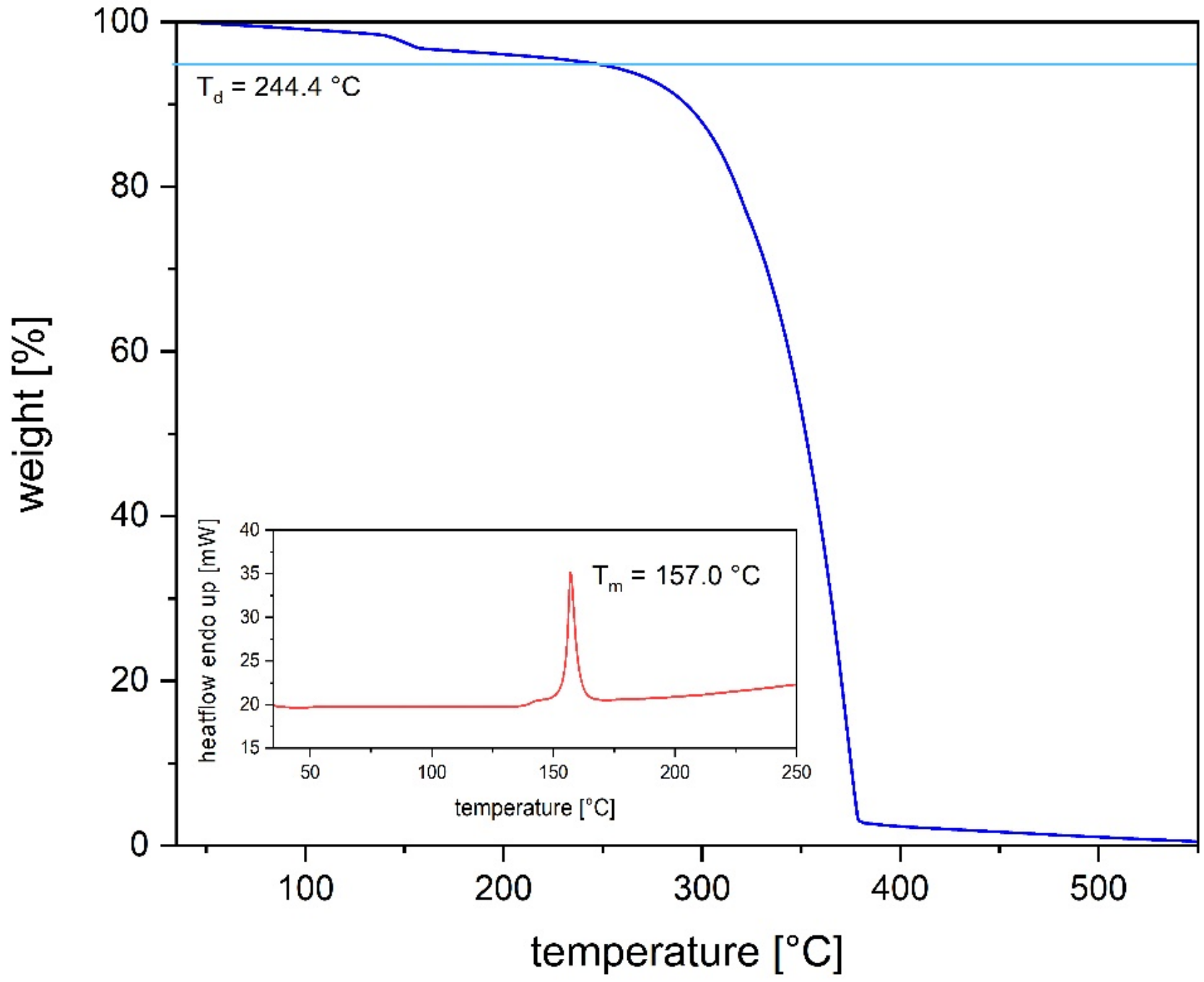

**Figure S18.** TGA and DSC (insert) results from simultaneous thermal analysis of **BP-PX** ($T_\mathrm{d}$ was obtained at 5% weight loss).

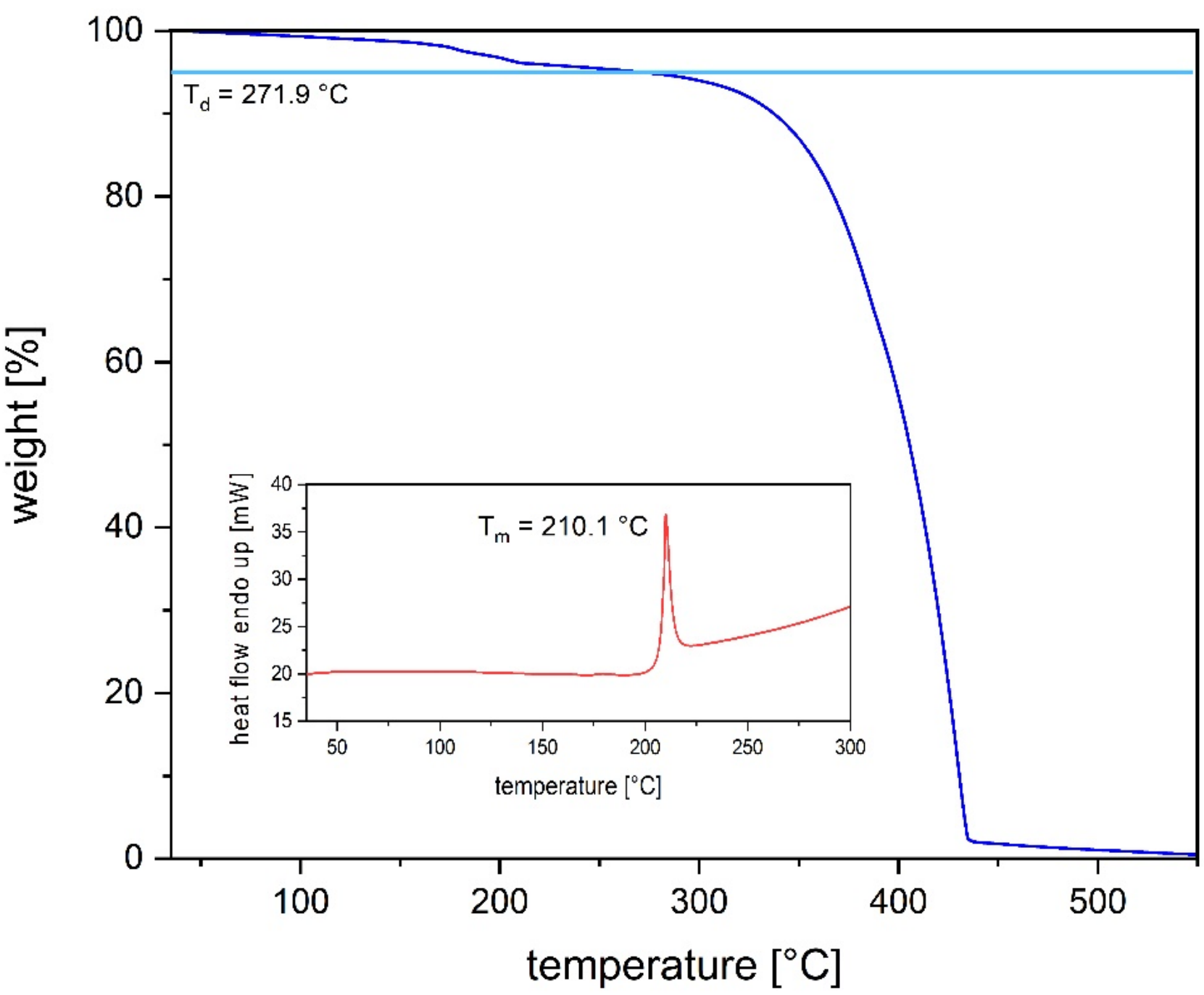

**Figure S19.** TGA and DSC (insert) results from simultaneous thermal analysis of **Py-2PX** ($T_\mathrm{d}$ was obtained at 5% weight loss).

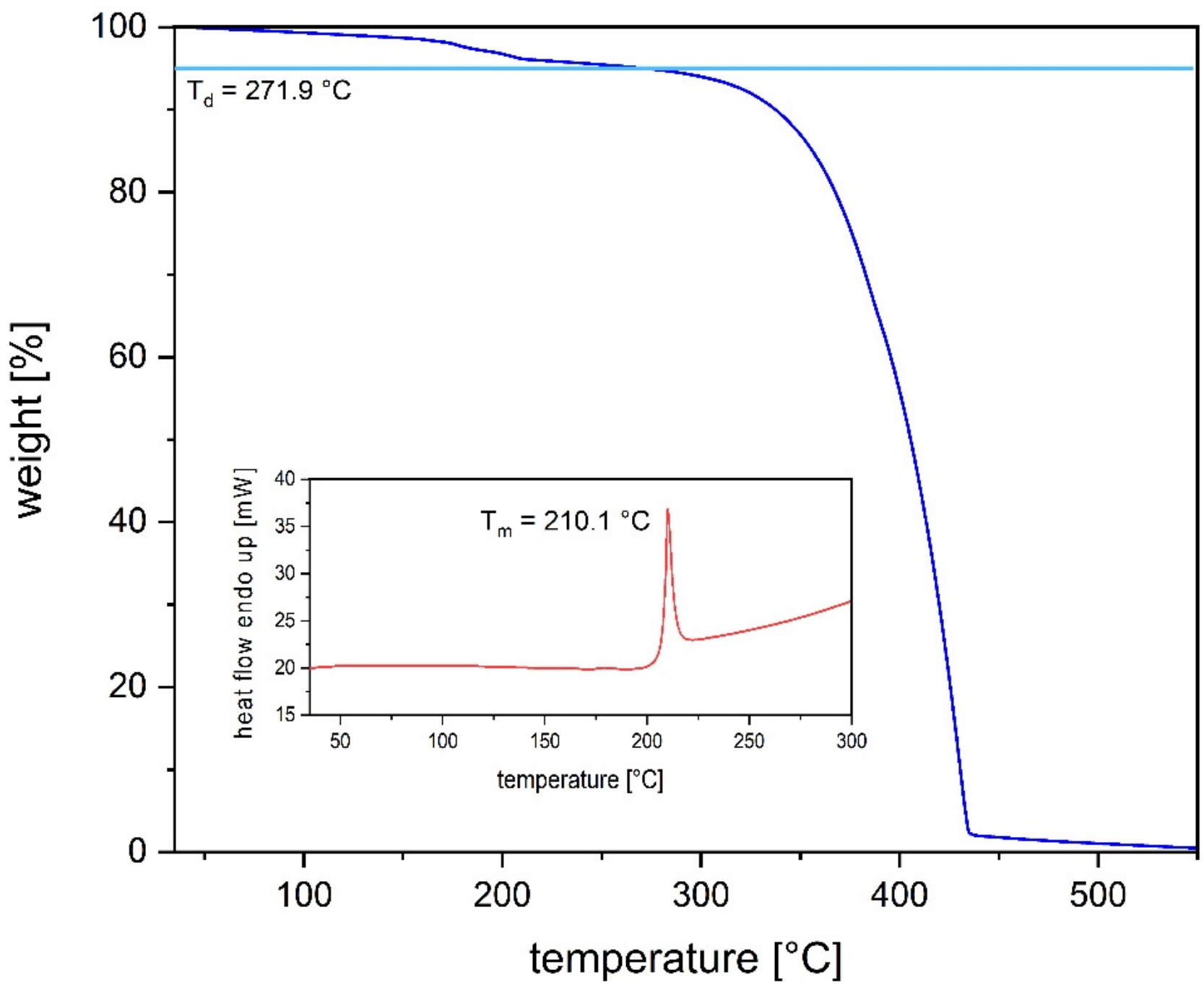

**Figure S20.** TGA and DSC (insert) results from simultaneous thermal analysis of **BP-2PX** ($T_\mathrm{d}$ was obtained at 5% weight loss).

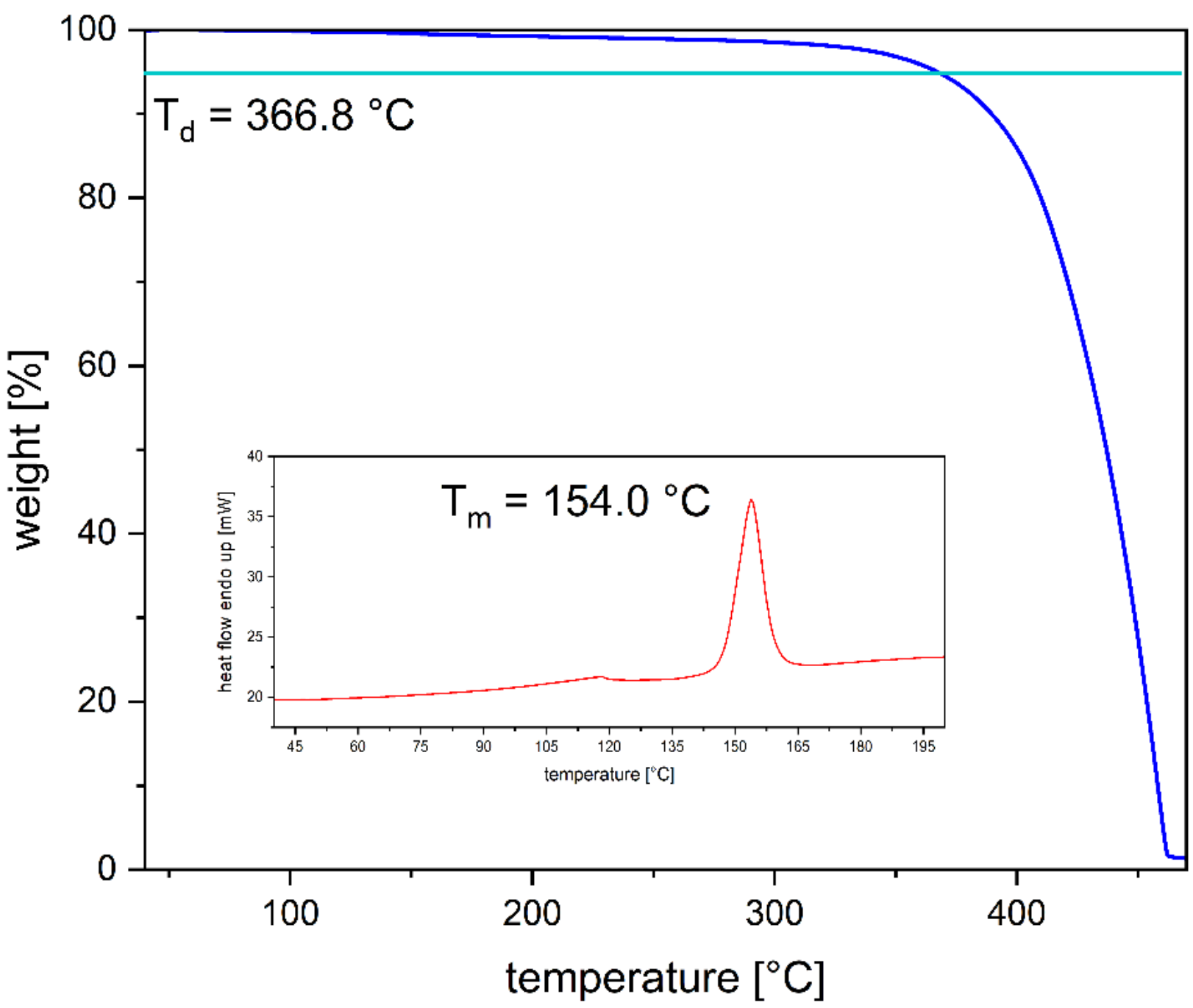

**Figure S21.** TGA and DSC (insert) results from simultaneous thermal analysis of **PX-Py-TA** ($T_d$ was obtained at 5% weight loss).

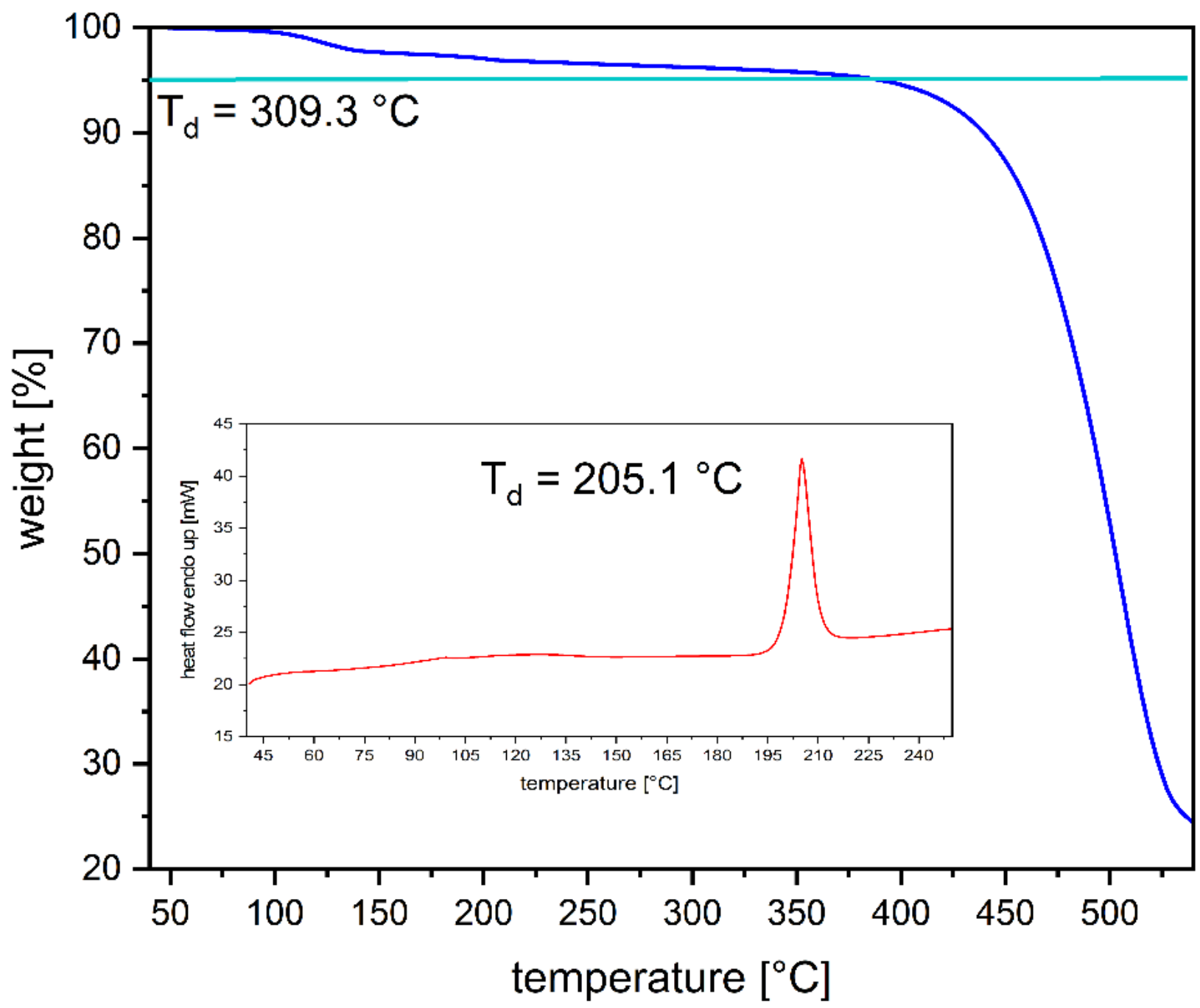

**Figure S22.** TGA and DSC (insert) results from simultaneous thermal analysis of **PX-BP-TA** ($T_d$ was obtained at 5% weight loss).

**S5 Further results from experimental photophysical characterization**

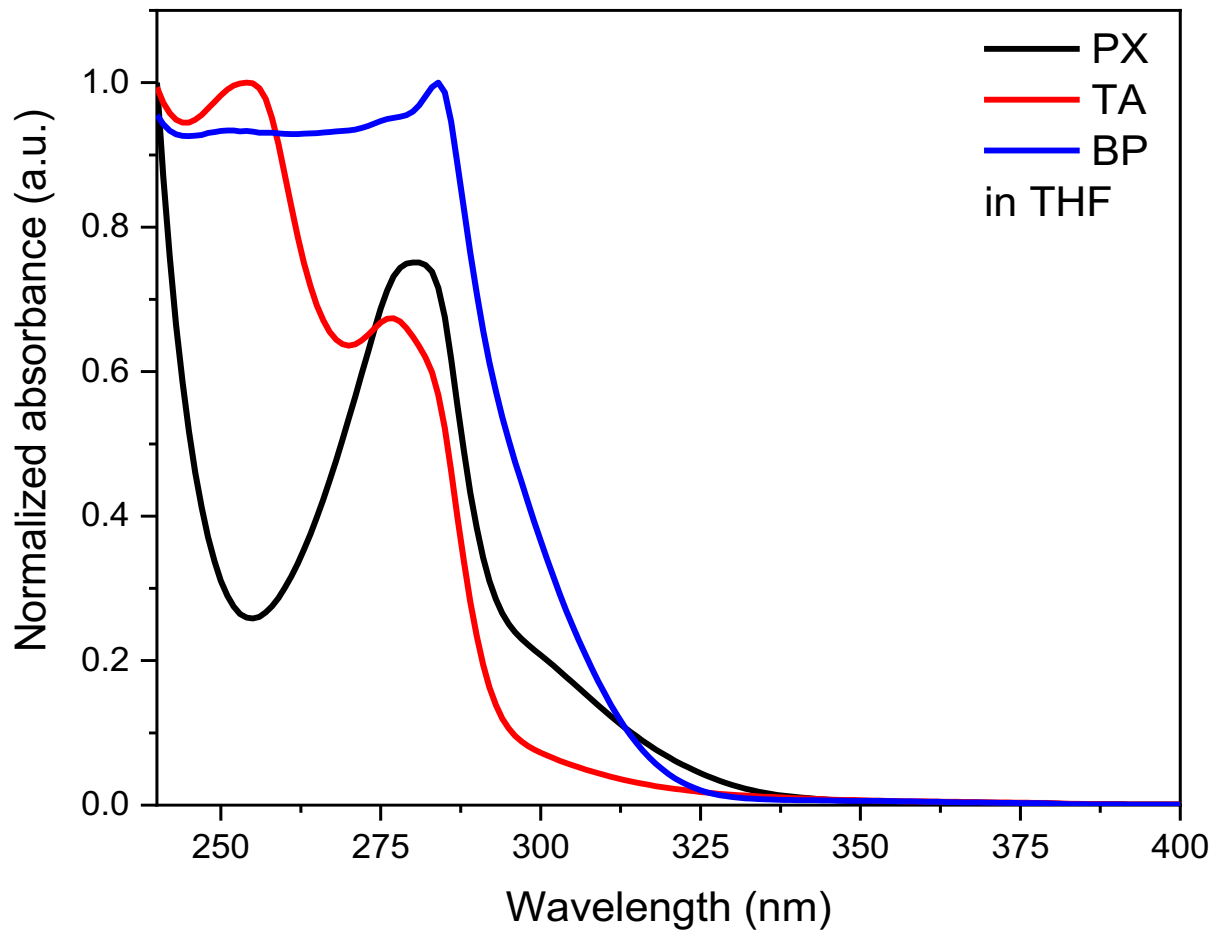

**Figure S23.** Absorption spectra of **PX**, **TA** and **BP** in THF.

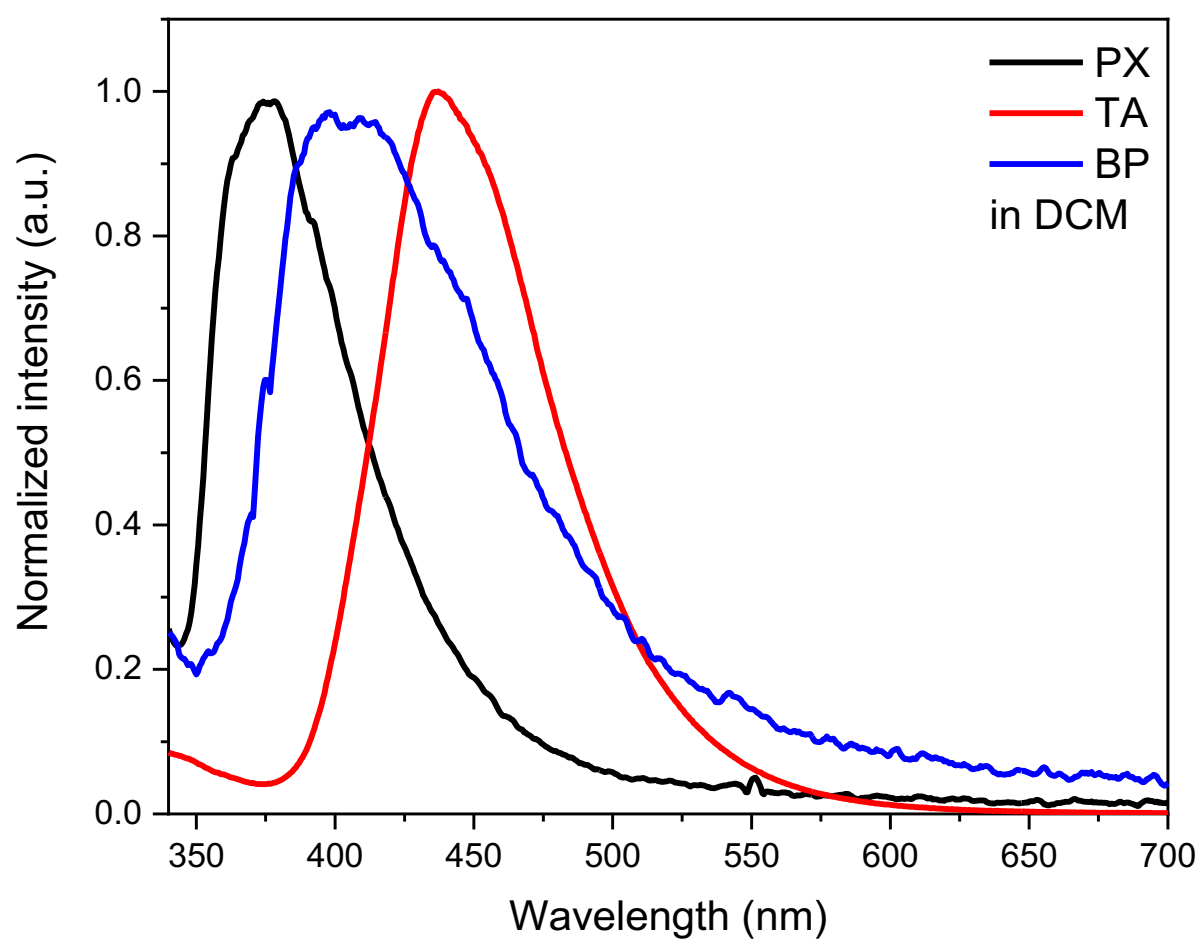

**Figure S24.** Emission spectra of **PX**, **TA** and **BP** in DCM with $\lambda_{\text{exc}} = 300$ nm at room temperature.

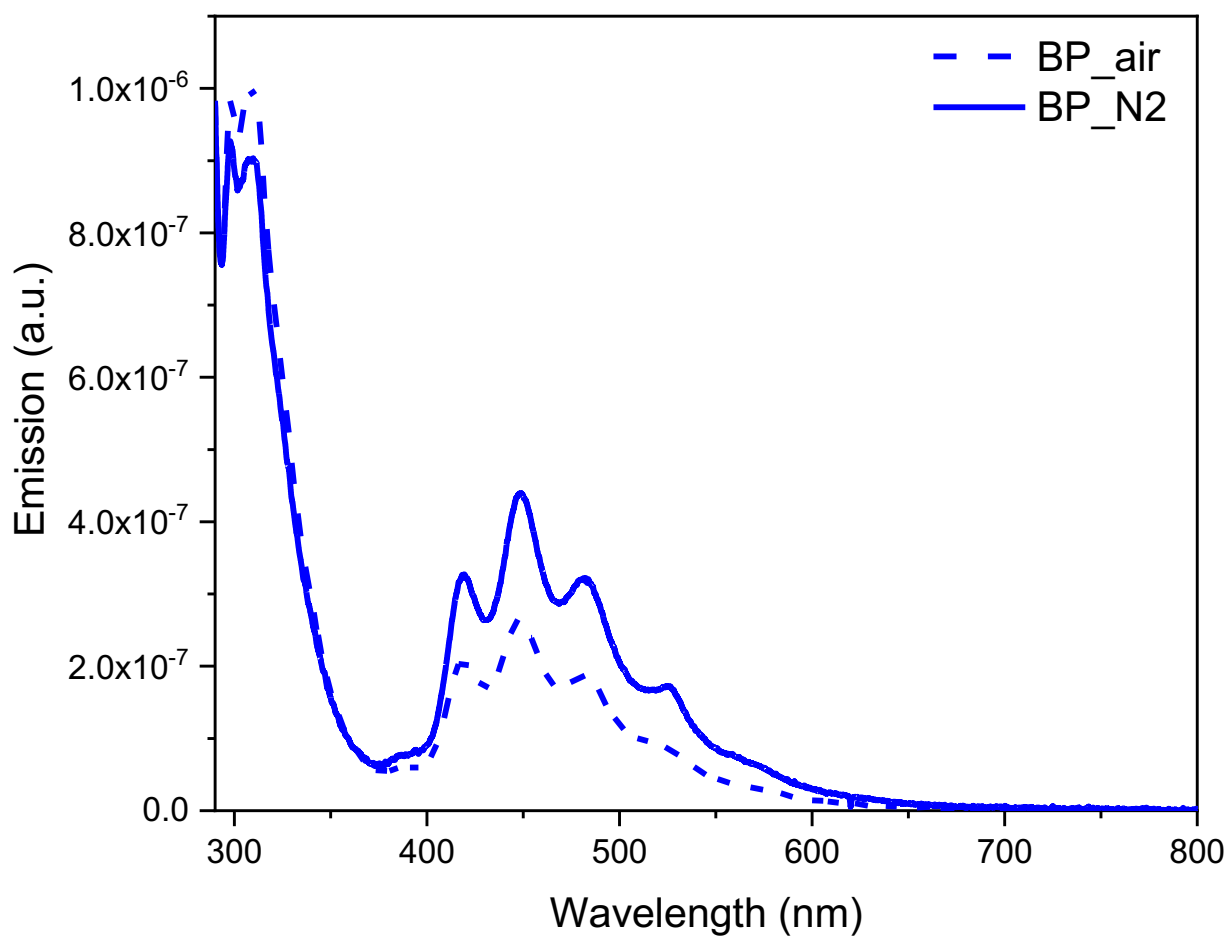

**Figure S25.** Emission spectra of **BP** in PS with $\lambda_{exc} = 275$ nm under aerated (dashed lines) and nitrogen atmosphere at room temperature.

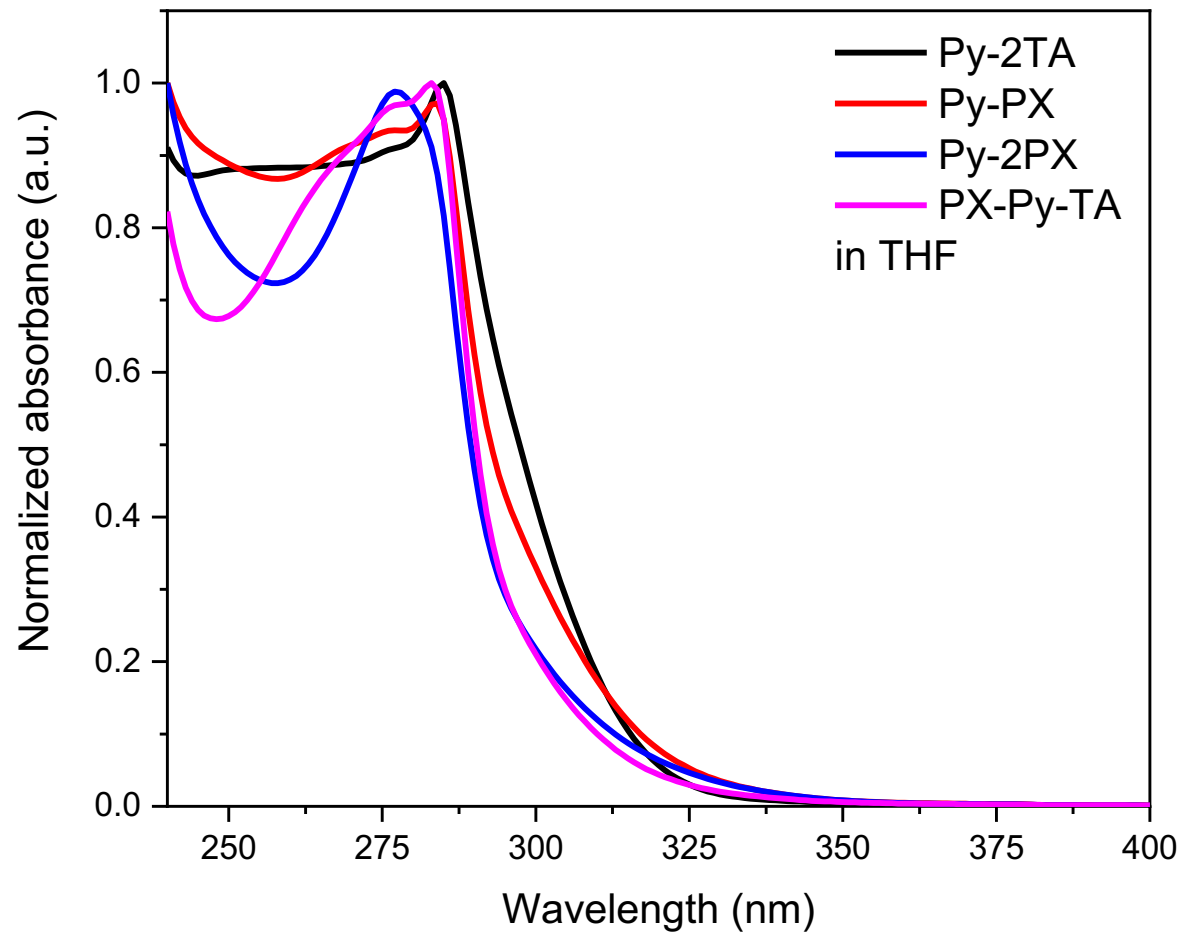

**Figure S26.** Absorption spectra of **Py-2TA**, **Py-PX**, **Py-2PX**, and **PX-Py-TA** in THF.

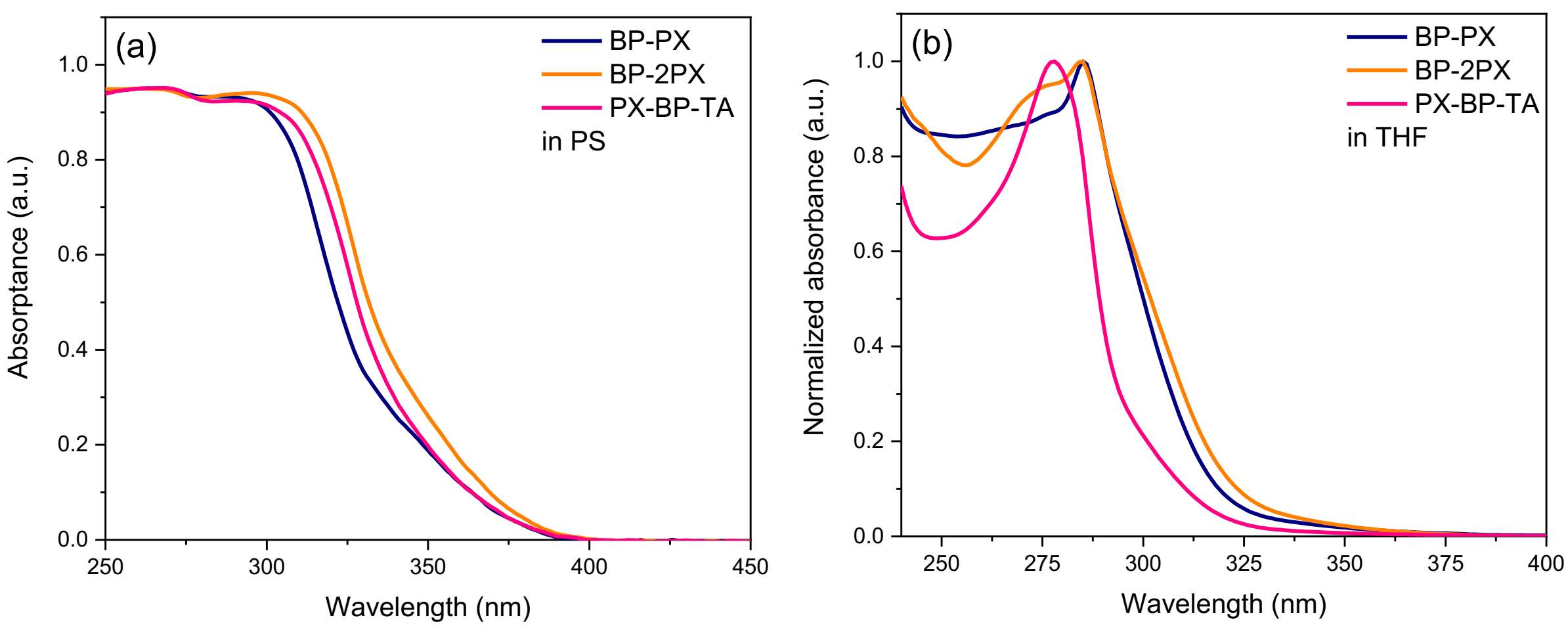

**Figure S27.** Absorptance spectra of **BP-PX**, **BP-2PX** and **PX-BP-TA** in PS (a), and absorption spectra in THF (b).

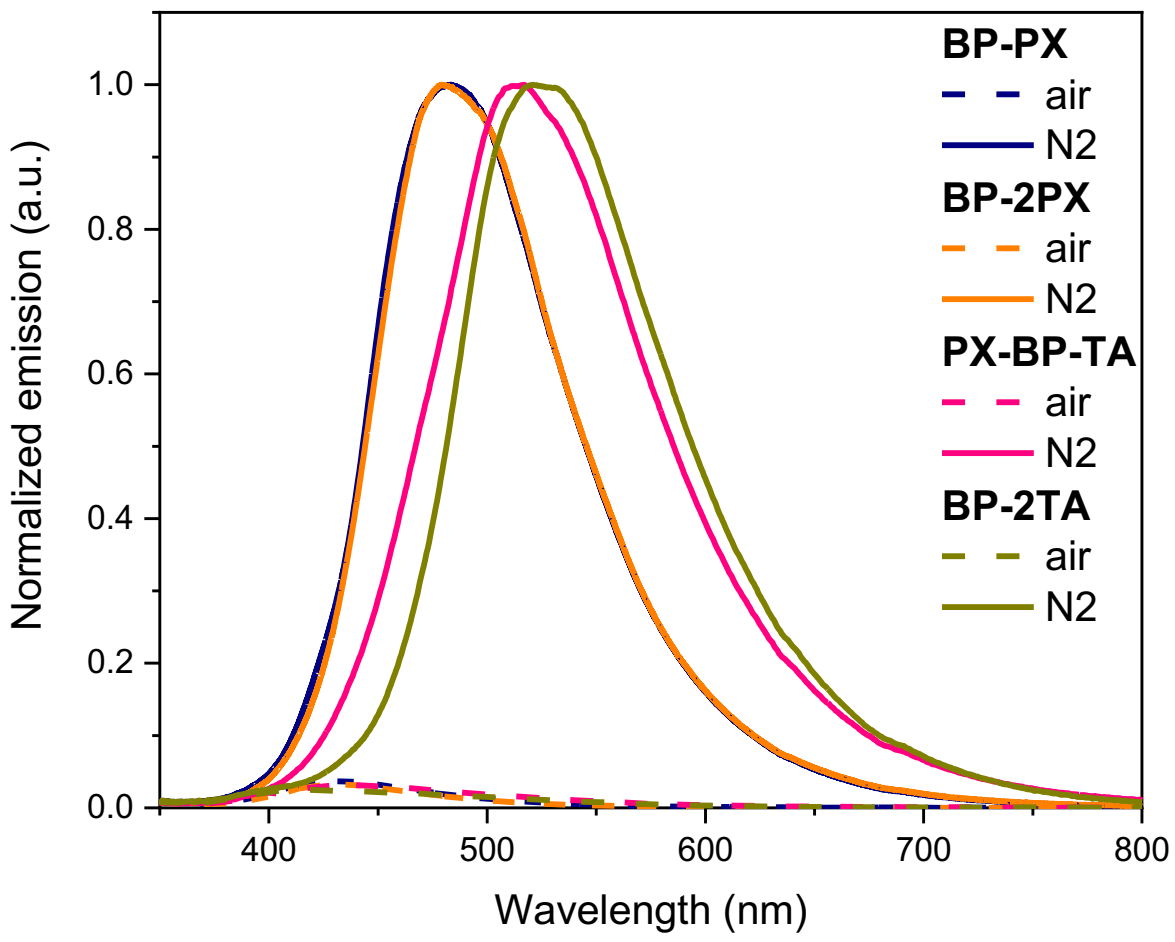

**Figure S28.** Emission spectra ($\lambda_{exc}$ = 275 nm) of **BP-PX**, **BP-2PX**, **PX-BP-TA** and **BP-2TA** in PS at room temperature under aerated (dashed lines) and nitrogen atmosphere.

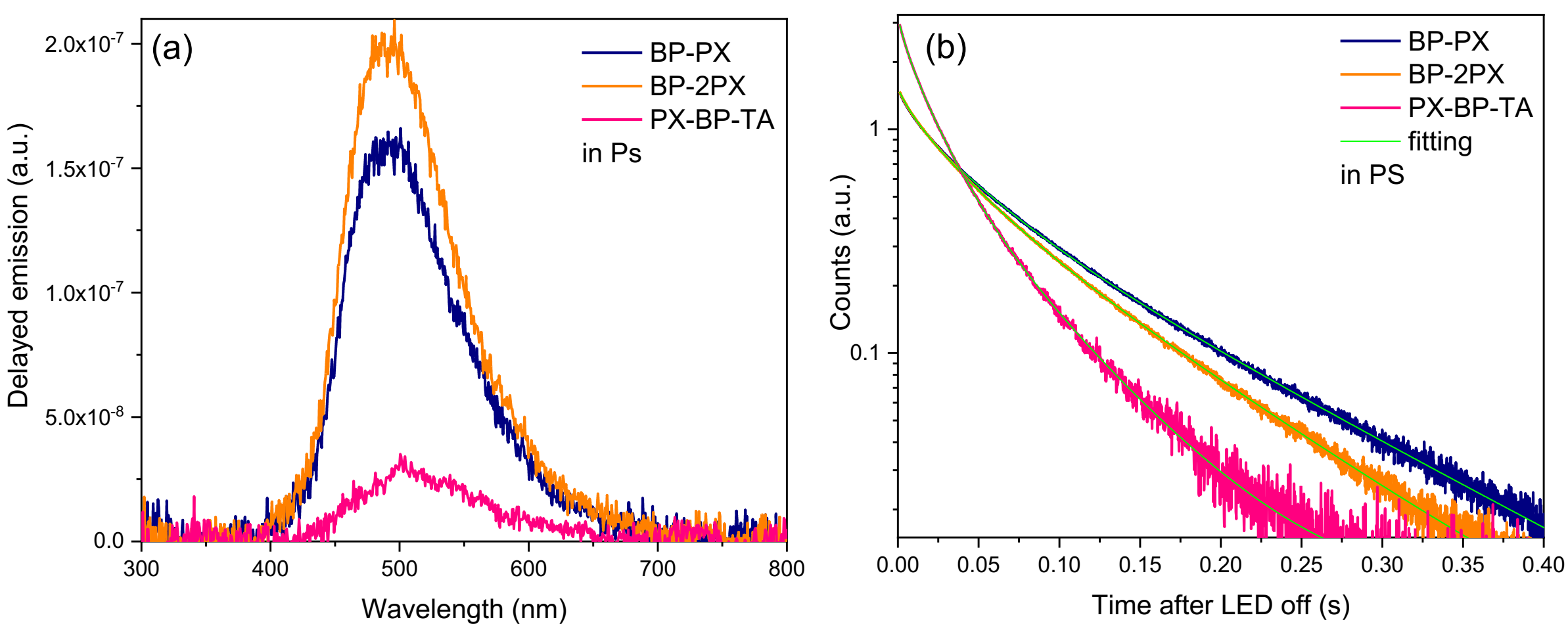

**Figure S29.** Persistent luminescence of the emitters. Delayed spectra (a) and corresponding phosphorescence decay (b) of **BP-PX**, **BP-2PX** and **PX-BP-TA** in PS at room temperature under nitrogen atmosphere collected at a delay time of 10 ms, showing only the phosphorescence ($\lambda_{exc}$ = 275 nm). Triexponential fit functions used to extract the phosphorescence lifetimes are presented as green lines in (b).

## S6 Characterization of programmable luminescent tags

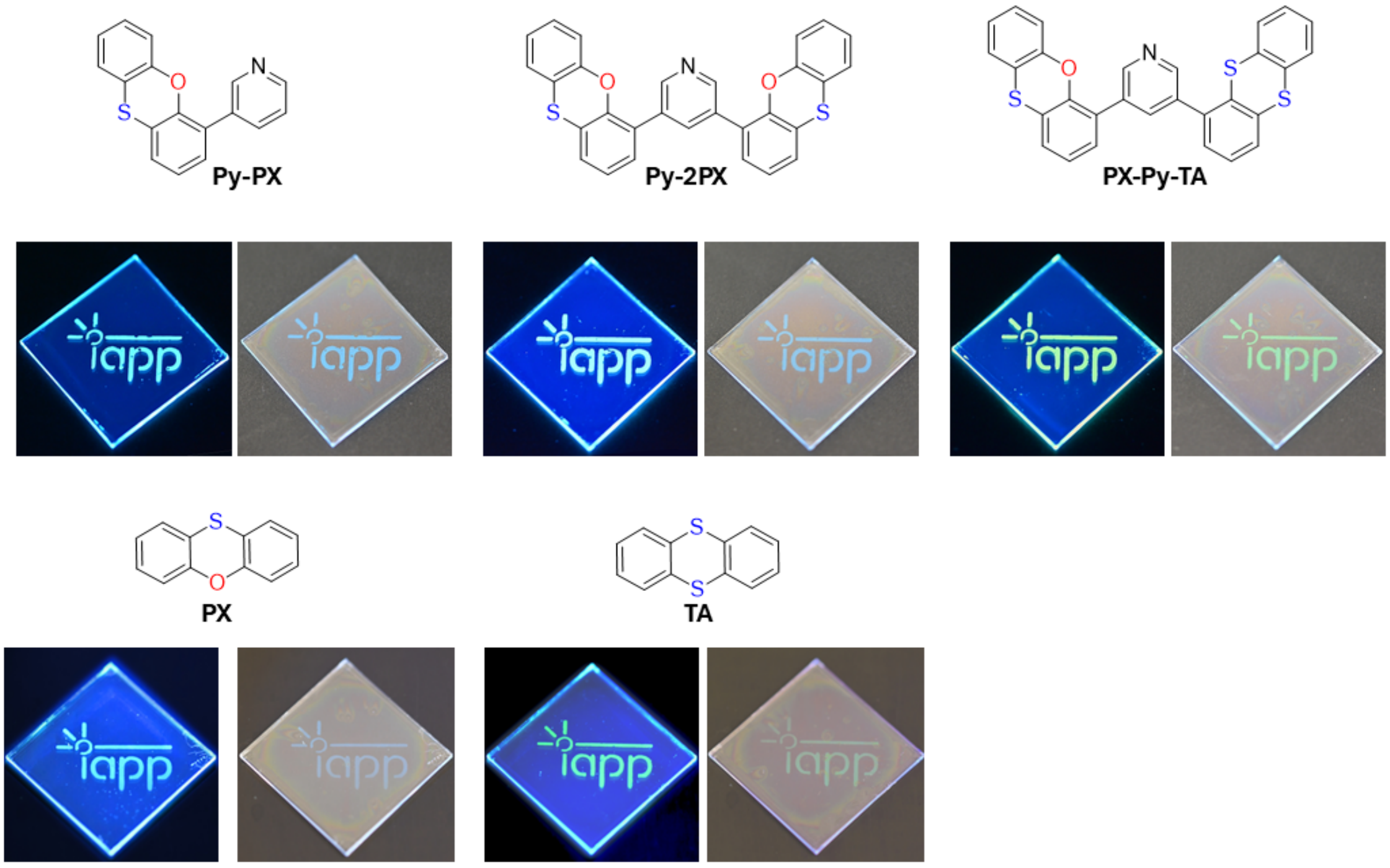

**Figure S30.** Representative illustration of information storage in PLTs using **Py-PX**, **Py-2PX**, and **PX-Py-TA**. All three compounds provide sufficient contrast even under regular lighting. The blue shift in phosphorescent emission for **Py-PX** and **Py-2PX** compared with **PX-Py-TA** is clearly visible